\documentclass[aip,jcp,twocolumn,groupedaddress]{revtex4}
\usepackage{epsfig}
\usepackage{color}
\usepackage{amsmath}

\begin{document}
\def\br{{\bf r}}
\def\vcmi{v_{cm,i}}

\title{A First Principle Approach to Rescale the Dynamics of Simulated Coarse-Grained Macromolecular Liquids}
\author{I. Y. Lyubimov}
\author{M. G. Guenza\footnote{Author to whom correspondence should be addressed. Electronic mail: mguenza@uoregon.edu}}
\affiliation{Department of Chemistry and Institute of Theoretical Science, University of Oregon, Eugene, Oregon 97403}
\date{\today}

\begin{abstract}
We present a detailed derivation and testing of our approach to rescale the dynamics of mesoscale simulations of coarse-grained polymer melts (I. Y. Lyubimov et al. J. Chem. Phys. \textbf{132}, 11876, 2010).  Starting from the first-principle Liouville equation and applying the Mori-Zwanzig projection operator technique, we derive the Generalized Langevin Equations (GLE) for the coarse-grained representations of the liquid. The chosen slow variables in the projection operators define the length scale of coarse graining. Each polymer is represented at two levels of coarse-graining: monomeric as a bead-and-spring model and molecular as a soft-colloid. In the long-time regime where the center-of-mass follows Brownian motion and the internal dynamics is completely relaxed, the two descriptions must be equivalent. By enforcing this formal relation we derive from the GLEs the analytical rescaling factors to be applied to dynamical data in the coarse-grained representation to recover the monomeric description. Change in entropy and change in friction are the two corrections to be accounted for to compensate the effects of coarse-graining on the polymer dynamics. The solution of the memory functions in the coarse-grained representations provides the dynamical rescaling of the friction coefficient. The calculation of the internal degrees of freedom provides the correction of the change in entropy due to coarse-graining. The resulting rescaling formalism is a function of the coarse-grained model and thermodynamic parameters of the system simulated. The rescaled dynamics obtained from mesoscale simulations of polyethylene, represented as soft colloidal particles, by applying our rescaling approach shows a good agreement with data of translational diffusion measured experimentally and from simulations.   The proposed method is used to predict self-diffusion coefficients of new polyethylene samples. 
\end{abstract}

\maketitle

\section{Introduction}
\label{SX:INTR}
\noindent
The past few years have witnessed a growing interest in the design and application of coarse-graining methods to simulate complex fluids.\cite{complex} This effort has been motivated by the need for  improving computational efficiency with the purpose of investigating complex systems on the numerous lengthscales on which their properties develop.\cite{Mueller-Plathe,DePablo,Klein,Klein1,Voth} Computer simulations have the capability of providing detailed microscopic information on the static and dynamics of the systems under study,\cite{Tisley} but they are limited in the range of timescales and in the number of molecules that can be simulated because the precision of the calculations degrades with the number of computer iterations with a behavior that depends on the Lyupanov exponent of the system. Once the number of particles is set, the window of achievable timescales that can be investigated becomes defined.\cite{Frenkel,shadowing} Because the maximum number of iterations decreases with increasing number of simulated particles, it is particularly difficult to simulate systems where characteristic lengthscales are diverging, such as a system approaching a second order phase transition.\cite{binderbook,blendjmg}

Recent improvements of computational machines has lead to a considerable extension of the maximum time- and length scales that can be reached by simulations where the system is described at the atomistic level. However, for many complex systems, including liquids of high-molecular weight macromolecules, the computational power is still inadequate to describe, at the atomistic level, the long-time dynamics.  For example, the most recent and advanced simulations of long chains that have an extended number of entanglements adopt a simplified model, which treats the structure of the polymer as a collection of beads and springs interacting through a FENE potential. This model allows for the simulations of a large number of polymers, which is important for the proper calculation of viscoelastic properties, and reaches full relaxation for all but the longest chains simulated.\cite{Grest}

Progress has been made when the focus is on qualitative behavior and scaling exponents.\cite{DPD,binderbook} For example, if the complex intra- and inter-molecular non-bonded interactions are simplified into an identical potential, the computational efficiency improves dramatically as the code does not need to identify and treat uniquely different pairs of interacting sites. This strategy, however, has the disadvantage that the thermodynamics of the system is not properly described  because the interactions are too drastically simplified. 

The need for methods that are fully predictive of the physical properties of a system on the basis of the specific chemical structure of the sample and its thermodynamic conditions has stimulated new interest  in developing fast quantitative simulations. Such predictive approaches are useful, for example, to evaluate \textit{a priori} the structure and dynamics of newly synthesized polymeric materials, in relation to their technological applications. Following this perspective, several procedures have been proposed to speed up atomistic simulations, while conserving their power of predicting quantitative properties.\cite{Mavran,Mavran1}  A few simulations of long entangled chains have been performed using united atoms (UA).\cite{Mavran,Putz,Tsolou,Harm}. For UA the effective unit is very close to the atom is size, i.e. $CH_x$ with $x=1,2,3$, which allows for some gain in the computational time.
  
A useful strategy to improve the outcome of simulations on the long timescale and large lengthscale is the use of coarse-graining procedures.\cite{MSMD, Mueller-Plathe} A coarse-graining procedure averages out irrelevant degrees of freedom, which occur on lengthscales smaller than a designated cutoff length, and this allows for the extension towards large scales of the simulations. Another way to put it is that, because the interaction potentials become softer,  the maximum time and  lengthscale increase as the basic timestep of the mesoscale (MS) simulations becomes larger. The characteristic lengthscale of coarse-graining has to be defined on the basis of the properties that need to be investigated. In this paper we discuss a first-principles way of selecting meaningful lengthscales for the structural and dynamical coarse-graining.

Several considerations need to be made to properly develop a coarse-graining procedure. As the coarse-grained liquid is represented as a function of new coordinates, an effective potential needs to be derived to be used as an input to the MS simulation. Care has to be taken to make the potential reproduce the structure of the system, namely pair distribution functions, and to be thermodynamically consistent.   A common procedure to optimize the coarse-grained description is to use self-consistent numerical methods that are optimized to reproduce atomistic descriptions through iterative procedures. Usually the target is the optimization of specific quantities, such as the pair distribution function,\cite{pdf} the forces generated by the soft potential,\cite{Voth} or directly  the thermodynamic properties.\cite{kroger} 

We recently proposed an approach that starts from the Ornstein-Zernike (OZ) equation where the atomistic sites are defined as real sites, and the coarse-grained sites are defined as auxiliary sites.\cite{YAPRL,melt,archi,blend,block,anthony,jaypaper}  Because our  procedure is analytical, and no optimization of parameters is needed in our approach as the potential is explicitly dependent on the thermodynamic and molecular parameters, it opens up the possibility of deriving a formal solution to key problems. For example, it is straightforward to show that the structural properties are consistent between the two levels of description, i.e. atomistic and coarse-grained.\cite{YAPRL} Moreover, the thermodynamic properties of the coarse-grained polymer liquid (e.g. isothermal compressibility,\cite{YAPRL} pressure in the virial and in the compressibility routes, total and cohesive energy) are shown to be formally consistent in the two levels of coarse-graining.\cite{mccarty} Local structure is easily included \textit{a posteriori} through a multiscale modeling procedure.\cite{jaypaper,multis} Finally, it is possible to derive an analytical rescaling factor for the dynamics, which is the main focus of this paper.

While the structure is well described by simulations of the coarse-grained system on the scale larger than the scale of coarse-graining,  the dynamics in MS simulations is unrealistically fast. Because local degrees of freedom are averaged out, the coarse-grained molecules move rapidly over a simplified free energy landscape. As the system explores efficiently this ``reduced" configurational landscape, the measured dynamics is artificially sped up by the smoothness of the potential. This is useful when coarse-grained representations are used to rapidly reach  an equilibrated state of the system before starting the atomistic molecular dynamics (MD) simulation. However, to directly collect information on the dynamics of systems from MS MD simulations, it is necessary to develop formalisms that rescale the unrealistically fast dynamics into the slower dynamics at atomistic resolution. In this paper we discuss in details an analytical procedure we recently proposed to rescale the mesoscale dynamics. The procedure is able to predict center-of-mass dynamics in quantitative agreement with experiments and atomistic simulations.\cite{shortivan}

The common strategy to rescale the dynamics  is to build a  ``calibration curve". The latter is obtained through the numerical fitting of dynamical quantities and optimization of the related parameters until the agreement of dynamical properties calculated in an all-atom and in a MS simulations is obtained.\cite{Nielsen,Kremer}  However, the numerical calculation of optimized calibration curves for the dynamics is quite difficult to achieve for macromolecular systems, as the dynamics is mode dependent: there are in principle $N$ internal modes in any molecule formed by $N$ units and the degree of polymerization of a long chain can be of the order of one million monomers. Moreover, numerically optimized parametric quantities are in general not transferable between systems in different thermodynamic conditions or with different chemical structure or increasing degree of polymerization. To overcome this problem, it is common to select as coarse-grained units ones that are very close in size to the atomistic units, so that the needed corrections to reach consistency in dynamic properties are minimal. In this case, corrections to the measured dynamics can be evaluated through a perturbative formalism which should rapidly converge to the desired value. The resulting computational gain is, however, still limited. Recently a numerical Ornstein-Zernike-based approach, with atomistic-level coarse-graining and the Percus -Yevick closure approximation, has been proposed, which shows different rescaling factors depending on the time correlation function under study.\cite{Wang} Another coarse-grained approach for polyethylene melts describes  a polymer chain as a collection of soft blobs connected by elastic bands, which enforce chain-chain uncrossability. Simulations follow an effective Langevin equation, whose parameters, i.e. effective potential, frictions and random forces, are obtained from an atomistic MD simulation. The optimized equation of motion (eom) reproduces well experimental data.\cite{PB}

Our approach is different from others in several ways. First of all it is analytical rather than numerical, providing the formal rescaling factor by solving the eoms in the two levels of representation. In this way, there is no need of performing an atomistic simulation to input numerical quantities in our formalism. The  dynamics measured in MS simulations of coarse-grained systems is directly rescaled into its atomistic counterpart using approximate closed-form expressions of friction and entropy. The two levels of coarse-graining, which allow for a straightforward analytical solution, are here two simple isotropic models: a soft sphere description for the molecular coarse-graining, and a bead-and-spring description for the monomer level coarse-graining. More sophisticated coarse-grained models can be developed for intermediate lengthscales\cite{block,anthony}; however, the formalism can become more involved.\cite{Akkerman} At the atomistic level the polymer is described as a "bead-and-spring" type of approach where the chain is a collection of friction points connected by harmonic springs. This is an implementation of the most popular model to treat unentangled polymer melt  dynamics, i.e. the Rouse model, and maps well into the dynamics of polymers described not only by UA simulations, but also by atomistic simulations and experiments, as it contains both local chemical structure, semiflexibility, and finite size effects.\cite{jpcmreview,blumen,Richter} It is a very accurate and molecular specific model, which has been shown to describe well, for example, the dynamics of the protein CheY, by testing its predictions of NMR relaxation against experiments.\cite{jpcmreview,Guenza}

In our coarse-grained model a polymer chain is represented as a soft-colloidal particle.\cite{Hall,Kremer1,Likos,YAPRL} Because the lengthscale of the coarse-graining is of the order of the molecular radius of gyration, i.e. the size of the molecule, the \textit{direct} predictions of the rescaling procedure are suitable for properties on lengthscales larger than $R_g$ and on timescales longer than the longest time of intramolecular relaxation, i.e. the longest correlation time in the Rouse theory. Internal dynamics cannot be obtained directly from the coarse-grained simulation; however, the rescaled diffusion coefficient leads to the monomer friction coefficient, which can be used as an input to well-tested theories of polymer dynamics, and \textit{indirectly} recovering the dynamics in the complete spectrum of polymer relaxation. An example of this kind of calculation is presented in this paper in Section \ref{SX:rft}. The extended level of coarse-graining provides a good test of our procedure, as large deviations could result from the rescaling if the method were not correct. Furthermore, our procedure can be useful in the study of long-time relaxation, given that large length scales and long timescales are most difficult to simulate for polymeric systems.

Although the outline of our rescaling theory has been published recently in a short paper,\cite{shortivan} this paper presents a  detailed derivation and discussion of our approach, which includes the prediction of the dynamics for new samples. After introducing our coarse-grained model input to the mesoscale simulations, we formally derive the rescaling approach for the dynamics, starting from the Liouville equation and using projection operators. Friction coefficients in the two descriptions are derived from the solution of the memory functions, while the rescaling of the simulation time is obtained from the entropic contribution, which accounts for the intramolecular degrees of freedom neglected in the soft colloid representation. Theoretical predictions compare well against UA MD simulations,\cite{MONDE,JARAM,HEINE,Mavran} and experiments,\cite{Richter,maybeothers,maybeothers1,maybeothers2,maybeothers3}. We also calculate the diffusion coefficient for new PE samples in thermodynamics conditions for which UA-MD data are not available. The purpose of  these calculations is to show that our method is not a simple rescaling of the mesoscale data through a shift of the diffusion coefficient to bring dynamical results to coincide with atomistic simulations, as is conventionally done. Instead our approach is fully predictive and can be used to calculated the diffusion coefficient, and the monomer friction, for new samples.

The paper is structured as follows: after introducing our coarse-grained model in Section \ref{SX:cg} and the projection operator technique to derive the equations of motion in the two levels of coarse graining in Section \ref{SX:dc}, we formally derive the rescaling approach for the dynamics from the solution of the memory functions in Section \ref{SX:ar}. We then present the MS simulations (Section \ref{SX:ms}), as well as the results obtained from the same,  and apply the rescaling procedure to the data from MS simulations (Section \ref{SX:ao}). Predictions of dynamical quantities and direct comparison for several samples, both from atomistic simulations and from experiments, provide a stringent test of the approach and show good quantitative agreement. A brief discussion in Section \ref{SX:da} concludes the paper.

\section{Coarse-graining of polymeric liquids: structural properties}
\label{SX:cg}
In this section we briefly review the theoretical background of the pair distribution functions that are input to our rescaling equation. The structure of a polymeric liquid, at lengthscales equal or larger than the monomer size, is fully specified by the momomer total distribution function, $h(r)$, which for polymer melts depends on two characteristic lengthscales, namely the density fluctuation lenthscale, which is the atomic length, and De Gennes' correlation hole lengthscale.\cite{GENNES} The latter is of the order of the molecular radius-of-gyration, $R_g=\sqrt{N/6}l $, which is the overall dimension of the polymer, where $N$ is the degree of polymerization and $l$ is the statistical bond length. We select $l$ and $R_g$ because these are the two lengthscales that define the structural properties of the polymeric liquid. 

\begin{figure}
\centering
\includegraphics[scale=0.5]{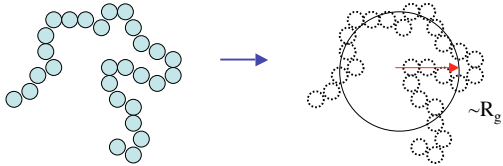}
\caption[]{Illustration of monomer and overall coarse-graining of a homopolymer linear chain.}
\label{models}
\end{figure}

At the monomer level traditional dynamical approaches, such as the Rouse model and semiflexible models, adopt a bead-and-spring representation where each monomer can be modeled as a friction point connected by springs (see Fig.~\ref{models}). A similar model, where the polymeric chain is described as a collection of  ``sites" centered at the center of the monomeric unit, is also in conventional theories of polymer liquids.\cite{PRISM,PRISM1} Although ``site" is the word most used in the liquid state community and ``monomer" or ``bead" is the common wording in the literature on polymer dynamics, in this paper they identify the same $CH_x$ unit and henceforth they will be used interchangeably. It is important to notice that all the $CH_2$ units are assumed to be equivalent and independent of the position along the chain.

The coarse-graining of a polymer at the $R_g$ lengthscale represents each molecule as an interacting soft colloidal particle with symmetric or asymmetric shape.\cite{Hall,Kremer1,Likos} In our model,\cite{YAPRL,melt,archi,blend} the  macromolecular liquid is represented as a liquid of symmetric soft colloidal particles interacting through a pair potential. This potential has a range of the order of few $R_g$, and each soft-colloidal particle is centered at the center-of-mass of a polymer (see Fig.~\ref{models}). 

The coarse-graining procedure that translates the monomer description into the solf-colloidal representation is performed starting from an Ornstein-Zernike equation where monomers are assumed to be real sites, while the center-of-mass (cm) are auxiliary sites.\cite{Hansen} The cm-cm total intermolecular correlation function is expressed as a function of the polymer parameters as \cite{YAPRL,melt} 
\begin{eqnarray}
\label{EQ:1}
h^{cc}(r) & = & \frac{3}{4} \sqrt{\frac{3}{\pi}} \frac{\xi'_\rho}{R_g} \left( 1-\frac{ \xi^2_c}{\xi^2_\rho} \right) e^{-3r^2/(4R^2_g)} \nonumber \\
&& -\frac{1}{2} \frac{\xi'_\rho}{r} \left( 1-\frac{\xi^2_c}{\xi^2_\rho} \right)^2 e^{R^2_g/(3\xi^2_\rho)} \Biggl[ e^{r/\xi_\rho} \mbox{erfc} \Biggl( \frac{R_g}{ \sqrt{3} \xi_\rho} +  \nonumber \\
&& \frac{ \sqrt{3}r}{2R_g} \Biggr) -e^{-r/\xi_\rho} \mbox{erfc} \left( \frac{R_g}{ \sqrt{3} \xi_\rho}- \frac{\sqrt{3}r}{2R_g} \right)\Biggr] \ , 
\end{eqnarray}
where erfc$(x)$ is the complementary error function. Here $\xi'_\rho=R_g/(2\pi\rho^*_{ch})=3/(\pi \rho l^2)$ with $\rho^*_{ch}\equiv\rho_{ch}R^3_g$ being the reduced molecular number density, $\rho_{ch}=\rho /N$ is the molecular density, $\rho$ the site number density, and $l$ is the statistical segment length. The length scale of density fluctuations, $\xi_\rho$, is  defined as $\xi^{-1}_\rho=\xi^{-1}_c +\xi'^{-1}_\rho$, and $\xi_c=R_g/\sqrt{2}$ is the length scale of the correlation hole.\cite{GENNES}

In the limit of long chains, $N\rightarrow \infty$, Eq.(\ref{EQ:1}) reduces to
\begin{equation}
h^{cc}(r)\approx-\frac{39}{16}\sqrt{\frac{3}{\pi}}\frac{\xi_\rho}{R_g}\left(1+\sqrt{2}\frac{\xi_\rho}{R_g}\right) 
\left(1 - \frac{9 r^2}{26 R_g^2} \right) e^{-\frac{3 r^2}{4 R_g^2}} \, .
\label{EQ:2}
\end{equation}
For polymers with $N \geq 30$, a plot of $h(r)$ shows that the two equations, Eqs.(\ref{EQ:1}) and (\ref{EQ:2}), are indistinguishable.\cite{YAPRL,melt} 

The structure of the liquid on the lengthscale of the polymer radius-of-gyration and larger, as represented by  $h^{cc}(r)$, is in quantitative agreement with the output of both the atomistic UA MD and the MS MD simulation of the coarse-grained liquid. The theory recovers identical analytical expressions of the compressibility in the atomistic and the coarse-grained representations, indicating thermodynamic consistency between the two levels of description.\cite{YAPRL,melt} 

Eqs.(\ref{EQ:1}) and (\ref{EQ:2}) are \textit{de facto} coarse-graining equations, which translate the atomistic description of a polymer liquid, onto its representation as a liquid of interacting soft colloidal particles of size $R_g$. The advantage of our coarse-graining approach is that it is analytical and general as it applies to systems with different thermodynamic conditions, different degree of polymerization and different bond length.\cite{YAPRL,melt,archi,blend,block,anthony,jaypaper,mccarty,multis}

\section{Dynamical coarse-graining: from the Liouville to the Langevin equations}
\label{SX:dc}
\begin{figure}[]
\begin{center} 
 \centering 
 \scalebox{0.35}{\includegraphics{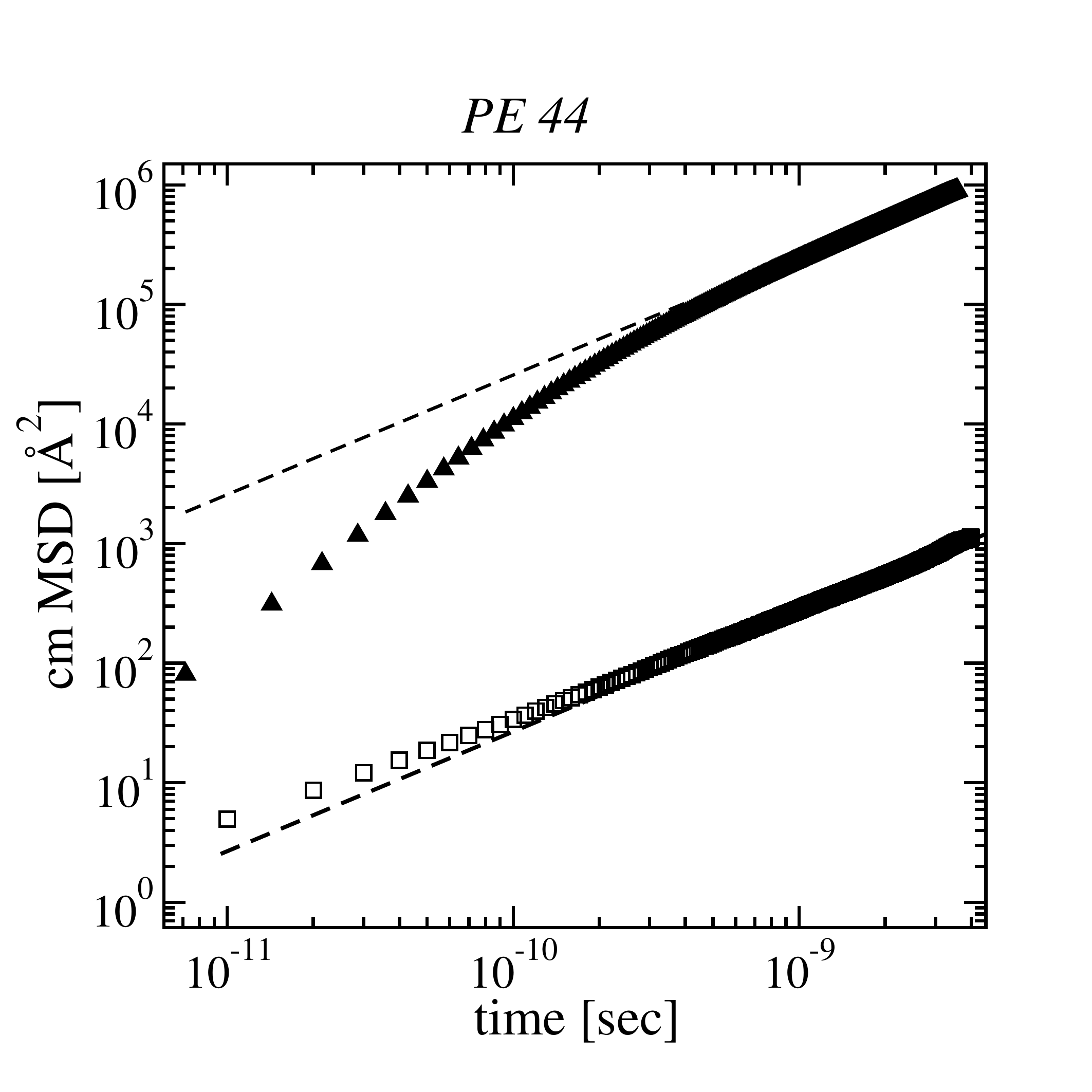}}
 \caption[Plot of the mean-square displacement for polyethylene melt 44]{Cm mean-square displacement, for a polyethylene melt with $N=44$, from MS MD simulations (triangles) and UA MD simulations (squares). Dashed lines show the diffusive limits of the two samples.}
 \end{center} 
 \label{figure2}
 \end{figure}

While the structure of the polymeric liquid, as represented by the total correlation function, is identical in the atomistic and coarse-grained descriptions,\cite{YAPRL,melt} the dynamics of the coarse-grained system, as measured in the MS MD simulations of the soft-colloidal particles, is unrealistically accelerated. In Fig.~(\ref{figure2}) we show, for a polyethylene chain with $N=44$, the mean-square-displacement of the center-of-mass  obtained in MS MD simulations of the polymer liquid represented as soft-colloidal particles and  the mean-square-displacement directly measured in UA MD simulations . The dynamics in the coarse-grained representation is several orders of magnitude faster than the atomistic description. Because the level of coarse-graining of the model presented here is extended, this effect is more evident than in other models; however, accelerated dynamics is present in any simulation of coarse-grained systems. 

It has been argued that there are two main effects of coarse-graining that accelerate the dynamics: namely, the change in entropy and the change in the friction coefficient. \"Ottinger has presented an approach for systems far from equilibrium that accounts for those effects.\cite{Ottinger} We propose here a procedure, based on first-principles theory to properly account for both contributions through the introduction of the necessary corrections for systems  where the fluctuation dissipation theorem applies, e.g. close to equilibrium. 
 
To coarse-grain the dynamics of the polymeric liquid on the lengthscale of the radius of gyration, we adopt a Mori-Zwanzig projection operator technique, where the selected slow variables are the position and momentum coordinates of the polymer center-of-mass. This description should represent well center-of-mass diffusion.\cite{SMPLQ,Ackasu,Opp}

The atomistic level representation is obtained following the same Mori-Zwanzing procedure, but choosing as the slow relevant variables the ensemble of position and momentum coordinates of the center of mass of the monomeric unit, which for a polyolefin is the $CH_x$ unit, with $x=1,2,$ or $3$. This model is consistent with the representation of the polyethylene chain in UA MD simulations,\cite{manychains,schweizer} and it has been shown to describe at a high level of accuracy the dynamics of polyolefins at the monomer lenghtscale.\cite{jpcmreview,Marina,marinamacrom}

In the long-time regime the two descriptions, soft-colloid and monomeric/UA, should be identical as they both recover the diffusive dynamics of the center-of-mass.\cite{DoiEdw} In fact, they are not, as the soft-colloidal description is heavily coarse-grained and its dynamics is accelerated. The analytical rescaling factor is derived directly from the comparison between the soft-colloid and the monomer dynamical equations.

As this coarse-graining and rescaling procedure is general, it can be adopted to formalize the dynamics of the molecular liquid at the desired level of coarse-graining. However, the projection operator technique rests on a separation of timescales between the slow relevant variables onto which the dynamics is projected, and the fast irrelevant variables that are averaged out. If no separation of timescales is observed, it is necessary to include corrections to the projected dynamics, which appear as contributions to the friction coefficient, expressed as memory functions. In the system investigated here, polymer melt dynamics, no clear separation of timescales occurs between the dynamics of the "tagged" chain and the dynamics of the surrounding molecules.\cite{manychains} For this reason, the Generalized Langevin Equation generated from this procedure needs to account for the correction terms to the projected dynamics, which are represented by the memory function contributions.\cite{Zwanzing}

For a liquid of  $n$ macromolecules containing $N$ monomers, the first-principle Liouville equation is simply written as
\begin{eqnarray}
\frac{\partial f \left( \mathbf{R, P}, t  \right)}{\partial t} = iLf \left( \mathbf{R, P}, t \right) \ ,
\label{EQ:Liouville}
\end{eqnarray}
with 
\begin{eqnarray}
 f \left( \mathbf{R, P}, t  \right)= \prod_{j=1}^{n} \left[ \prod_{a=1}^{N}  \delta (\mathbf{r}_a^j(t) - \mathbf{R}_a^j) \delta (\mathbf{p}_a^j(t) - \mathbf{P}_a^j)  \right] \ ,
\end{eqnarray}
the instantaneous distribution in reduced phase space, and $\mathbf{R}_a^i$ and $\mathbf{P}_a^i$ are the phase-space variables associated with the Cartesian position and momentum coordinates of the bead $a$ belonging to molecule $i$, namely $\mathbf{r}_a^i(t)$ and $\mathbf{p}_a^i(t)$.
The formal solution of Eq.(\ref{EQ:Liouville}) is 
\begin{eqnarray}
 f \left( \mathbf{R, P}, t  \right)= e^{-iLt}  f \left( \mathbf{R, P}  \right) \ ,
\end{eqnarray}
with the shorthand notation $ f \left( \mathbf{R, P}  \right) = f \left( \mathbf{R, P}, 0  \right) $

The Liouville operator is defined as 
\begin{eqnarray}
iL=- \sum_{j=1}^n \sum_{a=1}^N \left[\frac{\partial U_j}{\partial \textbf{r}_a^j} \cdot \frac{\partial}{\partial \textbf{p}_a^j} - \frac{\textbf{p}_a^j}{m}\cdot \frac{\partial}{\partial \textbf{r}_a^j}  \right] \ ,
\end{eqnarray} 
where the total energy $U_j$ in the Hamiltonian, $H$, contains both intramolecular, $U^0_j$, and intermolecular, $W_{ij}$, pairwise decomposable potential contributions. The intermolecular potential contains both interactions between the $n$ tagged chains, $W^0_{jk}$, and between the tagged chains and the surrounding ones, $W_{jk}$, so that the usual condition applies that $L_0f(\mathbf{R,P})=0$.
The statistical average of the phase space density is defined as 
\begin{eqnarray}
\Bigl< f \left( \textbf{R},\textbf{P} \right)\Bigr> = \int d \textbf{r} \int d \textbf{p}  f \left( \textbf{R},\textbf{P} \right) \psi \left( \textbf{r},\textbf{p}  \right) \ ,
\end{eqnarray}
with the equilibrium distribution of particle positions and coordinates 
\begin{eqnarray}
\psi \left( \textbf{r},\textbf{p}  \right) = e^{-\beta H} \left[ \int d \textbf{r} \int d \textbf{p}  e^{-\beta H} \right]^{-1} \ ,
\end{eqnarray}
where $\beta=(k_B T)^{-1}$, $k_B$ the Boltzman's constant, and $T$ the absolute temperature.
Following Mori-Zwanzig, we define the projection operator, $\hat{P}$, for the coarse-grained model we adopt, namely the monomer and the soft-colloidal. 

\subsection{Monomer level representation of the polymer chain}
\label{SX:ml}
In our atomic-level description each macromolecule is represented as a collection of connected beads, or friction points.  
In the field variables for one molecule ($n=1$), 
\begin{eqnarray}
 g \left( \mathbf{R, P}, t  \right)=  \left[ \prod_{a=1}^{N}  \delta \left( \mathbf{r}_a(t) - \mathbf{R}_a \right) \delta \left(\mathbf{p}_a(t) - \mathbf{P}_a \right)  \right] \ ,
\end{eqnarray}
the projection operator is defined as
\begin{eqnarray}
\hat{P}h \left( \textbf{R},\textbf{P},t  \right)  =  \int d \textbf{R}' \int d \textbf{P}' \int d \textbf{R}'' \int d \textbf{P}'' \nonumber \\
\bigl<h \left( \textbf{R},\textbf{P},t  \right) g \left( \textbf{R}',\textbf{P}'  \right)\bigr>  \nonumber \\
\times\bigl<g \left( \textbf{R}',\textbf{P}'  \right) g \left( \textbf{R}'',\textbf{P}''  \right)\bigr>^{-1}  g \left( \textbf{R}'',\textbf{P}'' \right) \ , 
\end{eqnarray}
where $\hat{P}=(\hat{P})^2$ and $\hat{P} g \left( \textbf{R},\textbf{P} \right)=g \left( \textbf{R},\textbf{P}  \right) $. Here we use for the field variable the symbol $g \left( \mathbf{R, P}, t  \right)$ to indicate that the slow variables in the projection operator can be different than the ones in the general formalism of the preceding section. By applying the projection operator to both the left and the right sides of the Liouville equation, one recovers a generalized Langevin equation.\cite{Ackasu,Opp,manychains,schweizer} 

Briefly, the generalized Langevin equation in the phase space is then transformed into its analog equation in space coordinates, yielding
\begin{eqnarray}
\label{inertiala}
m\frac{d^2 \textbf{r}_a(t)}{dt^2} & = &\beta^{-1} \frac{\partial}{\partial \textbf{r}_a(t)} \ln \psi(\textbf{r})\nonumber \\
&&  - \int_0^t d \tau  \sum_{b=1}^N \frac{\beta \textbf{p}_b}{3m}  \bigl<\textbf{F}_a(t) \cdot \textbf{F}_b^{\hat{Q}} (t-\tau)\bigr> \nonumber \\
&& + \, \textbf{F}_a^{\hat{Q}}(t) \ ,
\end{eqnarray}
where $\psi(\textbf{r}(t))$ is the intramolecular distribution function.
The inertial contribution in Eq.(\ref{inertiala}) can be discarded, as the liquid has a low Reynolds number and the dynamics is overdamped.
The Generalized Langevin Equation is simply written as
\begin{eqnarray}
\zeta_m \frac{d \textbf{r}_a(t)}{dt}=\frac{1}{\beta} \frac{\partial}{\partial \textbf{r}_a(t)} \ln \psi(\textbf{r}) +
 \textbf{F}_a^{\hat{Q}}(t) \ ,
\end{eqnarray}
with the averaged friction coefficient, in the Markov limit, 
\begin{eqnarray}
\zeta_m \approx \beta /3 \  N^{-1} \sum_{a,b=1}^N  \int_0^\infty  d \tau \bigl<\textbf{F}_a(t) \cdot \textbf{F}_b^{\hat{Q}} (t-\tau)\bigr>  \  .
\label{cJ}
\end{eqnarray}
This equation describes how the monomer friction coefficient is generated from the space and time correlation of the random forces that act on two different segments of the "tagged" polymer chain, $a$ and $b$. The extent of the correlation depends on the propagation of the forces through the macromolecule, its structure and local flexibility. The forces are generated by the monomers of the surrounding molecules randomly colliding with the monomers of the tagged chain: the collision strength depends on the structure of the liquid and on the interparticle potential.
A more explicit definition of the friction coefficient is given in the following sections.

\subsection{Solution of the Generalized Langevin Equation in the monomer representation}
\label{SX:so}
The intramolecular distribution function is approximated in our description by a Gaussian distribution
\begin{eqnarray}
\psi (\textbf{r})=\left[(2\pi)^N \det(\textbf{A}^{-1})  \right]^{-3/2} e^{-\frac{3}{2l^2}\textbf{r}^T\textbf{A}\textbf{r}} \ ,
\end{eqnarray}
which holds for polymer chains longer than about $30$ monomers.\cite{jpcmreview}
This leads to a  Generalized Langevin Equation where the intramolecular contribution is linear in the monomer coordinates 
\begin{eqnarray}
\zeta_m \frac{d \textbf{r}_a(t)}{dt} = -\frac{3k_B T}{l^2} \sum_{b=1}^N \textbf{A}_{a,b}\textbf{r}_b(t)+\textbf{F}^{\hat{Q}}_a(t) \  ,
\label{GLEmonomer}
\end{eqnarray}
and is simply solved through transformation into normal modes of motion.\cite{DoiEdw,jpcmreview}
The matrix $\textbf{A}$ is defined, for a semiflexible polymer represented as  a Freely Rotating Chain (FRC), as the product of two matrices, $\textbf{M}$ and $\textbf{U}$,
\begin{eqnarray}
\textbf{A}=\textbf{M}^T \left( \begin{array} {cc} 0 & 0 \\ 0 & \textbf{U}^{-1} \end{array} \right) \textbf{M} \ ,
\label{amatrix}
\end{eqnarray}
with the connectivity matrix, with dimensions $N \times N$, defined as
\begin{eqnarray}
\textbf{M}=  \left( \begin{array} {ccccc} N^{-1} & N^{-1} & N^{-1} & ... & N^{-1} \\ -1 & 1 & 0 & ... & 0 \\
0 & -1 & 1 & ... & 0  \\
... & ... & ... & ... & ... \\ 0 & ... & 0 & -1 & 1 \end{array} \right) \ ,
\end{eqnarray}
and the $\textbf{U}$ matrix defined as a function of the stiffness parameter $g$ as
\begin{eqnarray}
\textbf{U}_{ij}=\Biggl<\frac{\textbf{l}_i\cdot \textbf{l}_j}{|\textbf{l}_i| |\textbf{l}_j|}\Biggr>= g^{|j-i|} \ .
\label{semiflexibility}
\end{eqnarray}
Here, $g=-\bigl<\cos \theta\bigr>$ and $\theta$ is the angle between two consecutive bonds in the FRC representation of a homopolymer.\cite{DoiEdw}
The stiffness parameter, $g$,  is specific of the chemical structure and thermodynamic conditions of the sample under study.

\subsection{Center-of-mass level representation of the polymer chain}
\label{SX:co}
In the soft colloidal particle representation the projection operator targets the center-of-mass of the polymer. The field variable ($n=1$, $N=1$, $a=cm$) is  simply defined as
\begin{eqnarray}
 g \left( \mathbf{R, P}, t  \right)=  \left[  \delta \left( \mathbf{r}_{cm}(t) - \mathbf{R} \right) \delta \left(\mathbf{p}_{cm}(t) - \mathbf{P} \right)  \right] \ .
\end{eqnarray}
Applying the projection operator in the new field variable to the Liouville equation, where $U=0$ and $W_{ij}\neq 0$, leads to the generalized Langevin equation 
\begin{eqnarray} 
\frac{\partial}{\partial t} g \left( \textbf{R},\textbf{P},t  \right) & = &  -  \int_0^t ds \int d \textbf{R}' \int d \textbf{P}' M(\textbf{R}, \textbf{P},\textbf{R}',\textbf{P}') \nonumber \\
&& \times g \left( \textbf{R}',\textbf{P}',(t-s)  \right) + F \left( \textbf{R},\textbf{P},t  \right) \ , 
\end{eqnarray}
which reduces, following the procedure briefly outlined in Section \ref{SX:ml}, to
\begin{eqnarray}
\label{inertial}
m\frac{d^2 \textbf{r}_{cm}(t)}{dt^2} &=& -\int_0^t d \tau  \frac{\beta \textbf{p}_{cm}}{3m}  \bigl<\textbf{F}_{cm}(t) \cdot \textbf{F}_{cm}^{\hat{Q}} (t-\tau)\bigr> \nonumber \\
&& +\, \textbf{F}_{cm}^{\hat{Q}}(t) \ .
\end{eqnarray}
In the overdamped regime, 
\begin{eqnarray}
\zeta_{soft} \frac{d \textbf{r}_{cm}(t)}{dt} = \textbf{F}^{\hat{Q}}_{cm}(t) \  ,
\label{GLEsoft}
\end{eqnarray}
where $\zeta_{soft}$ is the friction coefficient for the colloidal particle, $\zeta_{soft}\cong  \beta /3 \int_0^\infty d \tau  \bigl<\textbf{F}_{cm}(t) \cdot \textbf{F}_{cm}^{\hat{Q}} (t-\tau)\bigr>$.
Eq.(\ref{GLEsoft}) obeys the fluctuation-dissipation relation $\left<\mathbf{F}_{cm}(t) \cdot \mathbf{F}_{cm}(t')\right>= \delta_{t-t'} 6 k_BT \zeta_{soft}$.

The choice of the field variables in the projection operator defines the length scale of coarse-graining and the variables in which the resulting Generalized Langevin Equation is expressed. Because the derivation just presented depends on the basic assumption that the correlation function of the bath variables  are short lived in the presence of heavy particles, and correction terms represented by the memory functions are minimized when a clear separation of timescale is observed between the slow variables in the projection operator and the fast variables that are averaged out, this criteria provides a way of selecting the relevant lengthscales for the  coarse-graining, when dynamical properties are under study. 

For example, as far as polymer dynamics is concerned, we know that for times longer than the longest Rouse correlation time, $\tau_R\approx R_g^2/D$, polymer internal dynamics is fully relaxed and the monomer dynamics follows the motion of the center-of-mass, which is long lived. This suggests that the center-of-mass coordinates are a good choice to represent the projected slow dynamics for time $t >> \tau_R$. This reasoning holds for both unentangled and entangled polymer dynamics as the longest relaxation time, after which free diffusion and Brownian motion set in, is $\tau_R$ with the proper diffusion coefficient, i.e. for unentangled chains $D_{unent}\propto N^{-1}$ and for entangled chains $D_{ent}\propto N^{-2}$.

\section{Analytical rescaling of the coarse-grained dynamics}
\label{SX:ar}
The two Langevin equations, Eqs.(\ref{GLEmonomer}) and (\ref{GLEsoft}), display the two levels of coarse-graining of the macromolecular liquid, which are adopted in this paper. The comparison of the two equations, which in the longtime regime should predict identical dynamics for the polymer center-of-mass, shows that the two equations differ because of the presence of the intramolecular free energy in the monomer description, which is absent in the soft colloidal approximation, and because of the different friction coefficients in the two representations. 

\subsection{Free energy rescaling}
\label{SX:fe}
The elimination of degrees of freedom increases the entropy of the system, as every coarse-grained state corresponds to a number of preaveraged microstates. In an extreme picture we can imagine that the preaveraging due to the coarse-graining procedure is in effect transforming the energy of the system, expressed for example in the Liouville equation by an Hamiltonian, into a free energy in the corresponding Langevin equation. While the Hamiltonian contains kinetics and potential energy, the free energy includes  an entropic contribution due to the preaveraged microstates for each coarse-grained state. 

As far as the free energy correction is concerned, the system described by the larger cutoff lenghtscale is the one where the level of coarse-graining is most extensive and the highest entropic correction has to be included.  This correction can be calculated from the comparison of the two equations. Because the system described at the monomer level is exploring in time the intramolecular energy states of the configurational landscape, its dynamics is slowed down with respect to the colloid representation where intramolecular degrees of freedom are not present. To take this effect into account we calculate the correction that has to be included in the soft-colloid representation to take into account the time spent by the atomic system to explore the internal degrees of freedom.

Consistent with the monomer-level model adopted in our study and with UA MD simulations, the polymer is described as a collections of beads, or friction points, connected by harmonic springs. Each bead corresponds to a $CH_x$ moiety, with $x=2$ or $3$, depending if the unit is imbedded in the chain or is terminal. This model has been shown to provide a realistic representation of the dynamics of numerous polymeric systems with different chemical structure.\cite{Guenza,jpcmreview,Marina,marinamacrom,blumen}

The intramolecular potential is defined as
\begin{eqnarray}
U(r)=\frac{3k_BT}{2l^2}\sum_{i,j=1}^N \textbf{A}_{i,j} \textbf{r}_i \cdot \textbf{r}_j \  ,
\label{EQ:3}
\end{eqnarray}
with $U(r)$ not to be confused with the semiflexibility matrix of Eq.(\ref{semiflexibility}). Here $\bf{A}$ is the connectivity matrix of Eq.(\ref{amatrix}), which represents the structure and local flexibility of the polymer \cite{Yamakawa,bixonzwanzig}, $\textbf{r}_i$ the position of bead $i$ in a chain of $N$ beads or united atoms, and $\textbf{l}_i=\textbf{r}_{i+1}-\textbf{r}_i$ the bond vector connecting two adjacent beads. 

The statistically averaged internal energy for one molecule consisting of $N$ monomers is given by
\begin{equation}
\left< \frac{U}{k_B T} \right> = N \int U e^{-\frac{3}{2 l^2} \mathbf{r}^T  \mathbf{A} \mathbf{r}} d\, r = \frac{3N}{2l^2} \int  \mathbf{r}^T  \mathbf{A} \mathbf{r} e^{-\frac{3}{2l^2} \mathbf{r}^T \mathbf{A} \mathbf{r}} \, .
\end{equation}
After solving the integral by normal mode transformation, as reported in Appendix I, this model predicts the average energy dissipated in the internal modes to be $\left< U/(k_B T) \right> = 3N/2$. The soft-colloidal representation, instead, has no internal degrees of freedom.

The simulation time $\tilde{t}$, as measured in the MS simulation of the coarse-grained system, translates into the real time $t$ after including the rescaling due to the energy, which is reduced by the amount of energy dissipated in the fluctuations due to internal degrees of freedom.\cite{Rice} For our model
\begin{eqnarray}
t=\tilde{t} R_g \sqrt{\frac{m}{k_BT} \frac{3}{2} N} \ ,
\label{EQ:4}
\end{eqnarray}
with the particle mass, $m$, and size $R_g$. 
This rescaling slows down the coarse-grained dynamics, but only partially accounts for the observed phenomenon because the rescaling of the friction needs to be included.

\subsection{Monomer friction coefficient}
\label{SX:mf}
The rescaling of the friction coefficient is calculated considering the friction of the polymer center-of-mass in the monomer/UA representation, and comparing the result with the friction of the cm of a soft colloidal particle. The expression for each of the friction coefficients is derived from its definition as the integral of the memory function contribution to the Generalized Langevin Equation (GLE) in the two levels of representation.

The effect of coarse-graining the Liouville equation, or projection onto the slow degree of freedoms, is the appearance in the Langevin equation of the dissipation terms, given by the random force and the friction coefficient. Systems with different levels of coarse graining have different friction and, as a consequence, different diffusion coefficients. 

For a particle in a liquid, the  center of mass mean-square displacement is defined as
\begin{equation}
\langle \Delta R^2(t) \rangle = 6\,D\,t \ ,
\end{equation}
with $D$ the diffusion coefficient. For a polymer, the cm diffusion coefficient is given by $D=k_BT/(N \zeta_{m})$, where $\zeta_m$ is the friction coefficient of a monomer, while for a liquid of soft colloidal particles $D_{soft}=k_BT/\zeta_{soft}$, with $\zeta_{soft}$ the friction coefficient of the colloidal particle. The two should be identical in the long-time limit, but they are not, as the diffusion coefficient obtained from MS MD simulation is much larger (much faster dynamics) than the one obtained from UA MD. The correction factor to scale down the MS MD diffusion coefficient, $D^{MS}$,  is 
 $\zeta_{soft}/(N\zeta_{m})$, which yields the rescaled mean-square displacement
\begin{equation}
\langle \Delta R^2(t) \rangle = 6\,D^{MS} \frac{\zeta_{\text{soft}}}{N \zeta_m}\,t \ .
\label{dmsris}
\end{equation}
The thermodynamic conditions of the system under study, i.e. density and temperature, and its molecular structure, i.e. the radius-of-gyration, enter the equation above both directly through the definitions of the friction coefficients, Eqs.(\ref{ciccio}) and (\ref{EQ:zeta_soft_mean_force}) and indirectly through the mesoscale simulation from which the diffusion coefficient, $D^{MS}$, is measured.

To solve Eq.(\ref{dmsris}) we start from the definition of  the monomer friction coefficient, $\zeta_m$, which is given  in the Markov limit by the memory function
\begin{eqnarray}
\zeta_{m} & \cong & \frac{1}{N} \sum_{a,b=1}^{N} \int_0^{\infty} d \tau \Gamma_{a,b} (\tau) \ .
\label{EQ:zetamo}
\end{eqnarray}
$\Gamma_{a,b}(t)$ is the function that
describes the correlation, through the polymer chain between monomers $a$ and $b$, of the random forces generated from the random collisions of the surrounding molecules undergoing Brownian motion, with \cite{Opp,manychains,schweizer}
\begin{multline}
\Gamma_{a,b}(t) \cong \frac{\beta}{3} \rho \int d\mathbf{r} 
\int d\mathbf{r'} g(r) g(r') F(r) F(r')\, \mathbf{\hat r} \cdot \mathbf{\hat r'} \\
\times\int d \mathbf{R} \, 
S^Q_{a,b}(R;t) S^Q(|\mathbf{r}-\mathbf{r'}+\mathbf{R}|;t) \, ,
\label{EQ:MFMON}
\end{multline}

\noindent where $g(r)=h(r)+1$ is the monomer radial distribution function, $F(r)$ is the total force exerted by all the matrix polymer on the monomer, and $S^Q(r;t)$ is the projected dynamic structure factor of the matrix fluid surrounding the polymer. The unit vectors  $\mathbf{\hat r}$ and $\mathbf{\hat r'}$ define the directions of the total exerted forces. The derivation of Eq.(\ref{EQ:MFMON}) is not completely new and is briefly reported in Appendix II. 

Eq.(\ref{EQ:MFMON}) rests on the approximations that the fluid is isotropic and that many-body correlation functions can be described with good accuracy as products of pair distribution functions. The solution of this equation is sometimes carried on by introducing a mode-coupling approximation,\cite{manychains,schweizer,fuchs} however we follow a different procedure. The dynamic structure factor, which is ruled by the projected dynamics, is approximated by its real dynamics counterpart, $S^Q(r;t) \approx S(r;t)$, simplifying the solution of Eq.(\ref{EQ:MFMON}). This is an acceptable approximation when the Langevin equation is expressed as a function of the slow variables\cite{Zwanzing} and holds for our system in the long-time, diffusive regime\cite{Marina}.

In order to separate the spatial coordinates of $S(|\mathbf{r}-\mathbf{r'}+\mathbf{R}|;t)$ in Eq.\ (\ref{EQ:MFMON}) it is convenient to use the Fourier transform
\begin{equation}
S(r;t)=\frac{1}{(2 \pi)^3} \int e^{i \mathbf{k r}} S(k;t) \, d \mathbf{k} \ ,
\label{EQ:FT}
\end{equation}
where the dynamic structure factor is calculated in reciprocal space as the sum of intra- and inter-molecular contributions
\begin{multline}
S(k,t) =\frac{1}{N}\sum_{\alpha\gamma}S_{\alpha\gamma}(k,t)=\frac{1}{N}\sum_{\alpha\gamma}\omega_{\alpha\gamma}(k,t)\\
+\rho \frac{1}{N}\sum_{\alpha\gamma}h_{\alpha\gamma}(k,t) \, .
\end{multline}

Here $\omega_{\alpha\gamma}(k,t)$ is the time dependent intramolecular  probability distribution functions for monomers $\alpha$ and $\gamma$, on the same molecule, to be separated by a reciprocal distance $k$, while $h_{\alpha\gamma}(k,t)$  is the corresponding intermolecular contribution.

Given that  the dynamics on the global scale is driven by the polymer diffusion, the intramolecular probability distribution function in reciprocal space can be expressed, in the limit of large lengthscales, $k \le 1/R_g$,  as
\begin{equation}
\omega_{\alpha\gamma}(k;t) \approx \exp\left[{-\frac{k^2 l^2|\alpha-\gamma|}
{6}}\right] \exp\left[{-k^2 D t}\right] \, ,
\end{equation}
where $D$ is the polymer center-of-mass diffusion coefficient and $l=N^{-1}\sum_{i=1}^N |{\bf{l}_i} |$ is the average segmental length. 

Because in Eqs.(\ref{EQ:zetamo}) and (\ref{EQ:MFMON}) the order of the summation and time integrals can be changed, the double summation reduces the  inter- and intramolecular distributions to their averages over the bead distribution. 
The site-averaged intramolecular probability distribution function, $\omega_0(k)$, is well approximated by the Debye formula,\cite{DoiEdw}
\begin{equation}
\omega_0(k) = \frac{1}{N}\sum_{\alpha\gamma}\omega_{\alpha\gamma}(k)=
\frac{2 N (e^{k^2 R_g^2}+k^2 R_g^2-1)}{k^4 R_g^4} \ ,
\label{EQ:Ave_wofk}
\end{equation}
\noindent or by its Pade' approximant
\begin{equation}
\omega_0(k) \approx \frac{N}{1+k^2 \xi_c^2} \ .
\label{EQ:PADE}
\end{equation}

The site-averaged intermolecular probability distribution is defined by the Ornstein-Zernike equation 
\begin{equation}
h(k)=\frac{1}{N}\sum_{\alpha\gamma}h_{\alpha\gamma}(k)=\frac{\omega_0^2(k)c(k)}{1-\rho c(k)\omega_0(k)} \, ,
\label{EQ:OZ}
\end{equation}
where $c(k)$ is the direct correlation function. At the monomer level we follow Curro and Schweizer's  PRISM thread approach,\cite{PRISM,PRISM1} where the polymer chain is modeled as a thread of vanishing thickness, $c(k) \approx c_0$, with $c_0=-(1-2 \xi_{\rho}^2/R_g^2)/(2 N^2 \rho_{ch} \xi_\rho^2/R_g^2)$. Substitution of $c_0$ and Eq.\ (\ref{EQ:PADE}) into Eq.\ (\ref{EQ:OZ}) gives 
\begin{equation}
h(k)=\frac{h_0}{(1+ k^2 \xi_\rho^2 )(1+ k^2 \xi_c^2)} \ ,
\label{EQ:OZnew}
\end{equation}
where $h_0=h(k=0)= (\xi_\rho^2/\xi_c^2 -1)/\rho_{ch}$ is related to the compressibility of the system.\cite{YAPRL,melt}

Because in the large length scale regime, of interest here, the relaxation of the liquid is dominated by the polymer diffusion, the dynamic structure factor is approximated as 
\begin{equation}
S(k;t)\approx S(k) \exp{\left[- k^2 D t \right] } \, .
\end{equation}

Finally, after introducing the integral representation of the delta function 
\begin{eqnarray}
\int d \mathbf{R} \ e^{i \mathbf{R}(\mathbf{k}_1+\mathbf{k}_2)}=
(2\pi)^3 \ \delta(\mathbf{k}_1+\mathbf{k}_2) \, ,
\end{eqnarray}
the last integral in Eq.(\ref{EQ:MFMON}) simplifies to
\begin{multline}
\int d \mathbf{R} \, S(R;t) S(|\mathbf{r}-\mathbf{r'}+\mathbf{R}|;t) =
\frac{1}{(2\pi)^6} \int d \mathbf{k}_1 \int d \mathbf{k}_2 \\
S(k_1;t) \ S(k_2;t) e^{i \mathbf{k}_2(\mathbf{r}-\mathbf{r'})}
\int d \mathbf{R} \ e^{i \mathbf{R}(\mathbf{k}_1+\mathbf{k}_2)} \ .
\end{multline}
Because the functions $\omega_0(k)$ and $h(k)$ are even with respect to $k$, the equation reduces to three contributions: the first is due to intramolecular interactions $\omega_0^2(k)$, the second includes the cross product $\omega_0(k) h(k)$, and the last is due to the intermolecular contribution $h^2(k)$.
This leads to the following expression 
\begin{multline}
\int d \mathbf{R} \, S(R;t) S(|\mathbf{r}-\mathbf{r'}+\mathbf{R}|;t) = 
\frac{1}{(2 \pi)^3} \int d \mathbf{k}\ \\
(\omega_0^2(k)+2 \rho h(k)\omega_0(k)+\rho^2 h^2(k)) e^{-2 k^2D t} e^{i \ \mathbf{k}(\mathbf{r}-\mathbf{r'})} \ .
\label{EQ:k_int_mon}
\end{multline}

Because we are assuming that monomers are interacting through a hard core potential, which is consistent with the PRISM thread model,\cite{PRISM,PRISM1} the force is a delta function and therefore
\begin{equation}
g(r)F(r)=g(d)\beta^{-1}\delta(r-d)\, .
\label{EQ:Hard_core}
\end{equation}
where $d$ is a hard core diameter, identical for any $CH_2$ bead in the chain, in the spirit of the UA-MD description and PRISM approach. When we compare our equations with data of experimental or simulated systems, where monomers interact through a Lennard-Jones potential, the latter has to be mapped onto a hard-core potential with the effective diameter, $d$.\cite{mcquarrie} 

The final expression for the monomer friction coefficient is given by
\begin{multline}
\zeta_m = \frac{1}{48\, \pi^3}\, \rho g^2(d)\, (\beta D)^{-1} \\
\times \left\{ J[\omega_0(k),\omega_0(k)]+2\rho J[\omega_0(k),h(k)]+\rho^2 J[h(k), h(k)] \right\} \ ,
\label{EQ:zetam}
\end{multline}
with the function
\begin{eqnarray}
\label{jei}
J[\alpha(k),\beta(k)] &=& \int d \mathbf{r} \int d \mathbf{r'} \int_0^\infty d k\, \frac{\sin(k |\mathbf{r}-\mathbf{r'}|)}{k|\mathbf{r}-\mathbf{r'}|}\, \mathbf{\hat r} \cdot \mathbf{\hat r'}\, \nonumber \\
&&\times \delta(r-d) \delta(r'-d)\, \alpha(k) \beta(k) \ .
\end{eqnarray}
The solution of Eqs.(\ref{EQ:zetam}) and (\ref{jei}) is given by a lengthy but analytical expression, which is a function of the molecular parameters, $\xi_{\rho}$, $R_g$, thermodynamic parameters, $\rho$, $\beta$, the diffusion coefficient, $D$, and of the hard-core diameter $d$, as

\begin{widetext}
\begin{eqnarray}
\label{ciccio}
\zeta_m & \approx & \frac{2}{3} (D\beta)^{-1} \rho g^2(d) \Biggl(\frac{1}{12}\pi N^2 d^2 R_g \Biggl[ 15\sqrt{2} + 40 \frac{d}{R_g} + 12\sqrt{2}\left(\frac{d}{R_g}\right)^2  \Biggr]+ \rho \pi N h_0 \frac{1}{3\sqrt{2}(R_g^2-2\xi_\rho^2)^2} \Biggl[ 12\sqrt{2}\xi_\rho^7 \\
&& + 12 d^4 R_g^3 \left(1-2\left(\frac{\xi_\rho}{R_g} \right)^2\right) +  \nonumber 4\sqrt{2} d^3 R_g^4 \left( 5-14 \left(\frac{\xi_\rho}{R_g}\right)^2 + 2 \left(\frac{\xi_\rho}{R_g}\right)^4 \right) + 3 d^2 R_g^5 \Biggl( 5- 14 \left(\frac{\xi_\rho}{R_g}\right)^2 - 4\sqrt{2} \left(\frac{\xi_\rho}{R_g} \right)^5 \Biggr)  \nonumber \\
&& - 12\sqrt{2} e^{-\frac{2d}{\xi_\rho}} \xi_\rho^7 \left(1+\frac{d}{\xi_\rho} \right)^2 \Biggr] + 
 \rho^2 \pi h_0^2 \frac{1}{12(R_g^2-2\xi_\rho^2)^3} \Bigg[ 40 d^3 R_g^6 + 15\sqrt{2} d^2 R_g^7 - 24\sqrt{2}d^4 R_g^3 \xi_\rho^2 - 144 d^3 R_g^4 \xi_\rho^2 \nonumber \\
&& +  6 \sqrt{2} d^4 R_g^5 \left(2 - 9 \left(\frac{\xi_\rho}{d}\right)^2 \right) +  12 R_g^2 \xi_\rho^7 \left(4 \left(\frac{d}{\xi_\rho}\right)^3 - 7 \left(\frac{d}{\xi_\rho}\right)^2 + 9 \right) - 8\xi_\rho^9 \left(4 \left(\frac{d}{\xi_\rho}\right)^3 - 9 \left(\frac{d}{\xi_\rho} \right)^2 + 15 \right) -  \nonumber\\
&& e^{-\frac{2d}{\xi_\rho}} 12 \xi_\rho^4 (d + \xi_\rho) \left( R_g^2 (d+3\xi_\rho)(2d+3\xi_\rho)-2\xi_\rho^2(2d^2+5\xi_\rho d + 5 \xi_\rho^2) \right) \Biggl] \Biggl) \  . \nonumber
\end{eqnarray}
\end{widetext}

This expression is general and holds for any homopolymer melt represented as a collection of identical beads interaction through a hard-core potential of range $d$. The value of $d$ is specific of the monomeric structure of the homopolymer.

\subsection{Friction coefficient for a liquid of interacting soft colloidal particles}
\label{SX:fc}
The friction coefficient for a point particle interacting through a soft repulsive potential is much smaller than the friction of the macromolecule before coarse-graining. In fact, the friction coefficient of an object can be estimated using Stokes' formula where $\zeta=6 \pi \eta r_H$, with $\eta$ the fluid viscosity and $r_H$ the hydrodynamic radius. The latter can be evaluated from the surface area of the object exposed to the solvent, which can be estimated by "rolling" a solvent molecule on the object. It is evident that the surface available to the solvent in a bead-spring representation of a polymer is much higher than the surface available to the solvent for a point particle interacting through a soft, long-ranged potential.

To calculate the friction coefficient for a soft colloidal particle, we start from the Generalized Langevin Equation that describes the time evolution  for the position coordinate of the molecular center-of-mass, i.e. Eq.(\ref{GLEsoft}) where the friction coefficient for soft particles is given by 
\begin{eqnarray}
\zeta_{\text{soft}}&\cong& (\beta/3) \  \rho_{ch} \int_0^{\infty}d\,t \  \int d\mathbf{r} \int d\mathbf{r'} g(r) g(r') F(r) F(r')\, \nonumber \\
&&\times\mathbf{\hat r} \cdot \mathbf{\hat r'} \int d \mathbf{R} \, S(R;t) S(|\mathbf{r}-\mathbf{r'}+\mathbf{R}|;t) \, .
\label{EQ:zeta_soft}
\end{eqnarray}

Eqs.(\ref{EQ:MFMON}) and (\ref{EQ:zeta_soft}) look identical, with just a different form of the density prefactor. In reality the form of the pair-distribution function, $g(r)$, the force exerted by the surrounding molecules on the tagged chain, $F(r)$, and the dynamic structure factors, $S(\mathbf{R},t)$, are different quantities in the monomer and soft-colloid representations.

We assume that the dynamic structure factor in reciprocal space has the form
\begin{equation}
S(k;t)\approx S(k)\ e^{-D t k^2}=(1+\rho_{ch} h^{cc}(k))\ e^{-D t k^2}
\label{EQ:SFSC}
\end{equation}
where $h^{cc}(k)$ is the center of mass total pair correlation function  
\begin{equation}
h^{cc}(k) = h_0 \left[\frac{1+k^2 R_g^2/2}{1+k^2 \xi_\rho^2} \right] e^{-\frac{k^2 R_g^2}{3}} \, .
\label{EQ:hcck_OZ}
\end{equation}
with $h_0=(\xi_\rho^2/\xi_c^2 -1)/\rho_{ch}$, as defined in the previous section.
 Eq.(\ref{EQ:hcck_OZ}) is just the Fourier transform of Eq.(\ref{EQ:1}).
 Eq.(\ref{EQ:SFSC}) indicates that in the long-time regime, which is of interest here, the relaxation of the liquid is largely driven by the center-of-mass diffusion, while internal dynamics and local modes of motion are already fully relaxed. This is a reasonable assumption given that the lengthscale of our treatment is the overall polymer dimension, and no structural or dynamical information is retained on the local scale.

To perform our calculation we need to defined an approximate analytical form of the effective force. To do so we adopt the simplified form of $h^{cc}(r) $, Eq.(\ref{EQ:2}). Then, we reduce further the expression by neglecting the small attractive component of the potential. Finally we approximated the real potential, $v(r)$, with its mean-force counterpart $w(r)\approx - k_B T \ln{[h(r)+1]}$ properly rescaled. The real potential, calculated through the HNC approximation as described in the following section, is a complicated function of $h(r)$. However it can be related, in an approximated way, to the simpler potential of mean force through the equation
\begin{equation}
v(r)\approx \frac{v(0)}{w(0)}w(r) \ ,
\end{equation}
where $v(r) \approx \sqrt{3} w(r)$ for all the samples considered in this study.
These approximations define the force $F(r)$ and the pair distribution function, $g(r)=h(r)+1$,  entering the equation for the friction coefficient.

The resulting expression for the friction coefficient of the soft colloidal particle is expressed as a function of the diffusion coefficient $D$, $\beta$, $\rho_{ch}$ and the two length scales $R_g$ and $\xi_\rho$ as
\begin{eqnarray}
\zeta_\text{soft} &\cong& \frac{4}{3} \sqrt{\pi} (D \beta)^{-1} \rho_{ch} R_g \xi_\rho^2 \left(1+\frac{\sqrt{2}\xi_\rho}{R_g} \right)^2 \frac{507}{512} \nonumber\\
&&\times\left[ \sqrt{\frac{3}{2}}  + \frac{1183}{507} \rho_{ch} h_0 + \frac{679\sqrt{3}}{1024} \rho_{ch}^2 h_0^2 \right] \, .
\label{EQ:zeta_soft_mean_force}
\end{eqnarray}
This expression is an approximated analytical form for the friction coefficient of a soft colloidal particle.

\section{Mesoscale Simulations}
\label{SX:ms}
Here we present numerical calculations to illustrate and discuss the rescaling procedure of the preceding sections. We first perform MS MD simulations of the coarse-grained polymer liquid, where each chain is represented as a soft colloidal particle, centered at the center-of-mass of a chain, and interacting with the surrounding particles through a soft repulsive potential of the order of few times the chain dimension, $R_g$. The simulations of the soft colloidal liquid produce dynamical properties that are accelerated due to the soft nature of the potential in the coarse-grained representation. These properties are rescaled following our procedure, and then compared with existing data, when they are available.

In a previous paper we briefly presented calculations of the rescaled dynamics for a variety of systems including UA MD  simulations and experimental data of PE diffusion available in the literature, that we use to test the accuracy of our procedure. We selected UA MD simulations as our test  (see Table \ref{TB:parameters_UAMDa}) because they have been shown to reproduce with a high level of accuracy the dynamical properties  of PE melts, such as diffusion and viscosity.\cite{PaulSmith,PaulSmith1} We also compared predictions of rescaled MS-MD simulations with experiments for samples with temperature $T=509 \ K$, monomer site density $\rho =0.0315302$~sites/\AA$^3$  and $N=  36, \ 72, \ 106, \  130, \ 143, \ 192,$ and $ 242 $. \cite{Richter,maybeothers,maybeothers1,maybeothers2,maybeothers3}. Our MS-MD simulations, properly rescaled, provided good quantitative predictions of the diffusion coefficient for those systems.\cite{shortivan}

In this paper we use those same systems to illustrate our procedure. Moreover we present new results for PE samples, not present in the literature, to underline the predictive power of the theory, where no calibration curve is necessary. Once a system is selected, its structural and thermodynamic parameters are defined  and are used as input to the MS MD simulation so that the whole procedure is free of adjustable parameters, with the exception of the parameter $d$ that is fixed for PE once and for all samples.\cite{MONDE,JARAM,HEINE,Mavran}  

Systems that we simulated include liquids of chains with increasing degree of polymerization, as described above.
As the molecular weight of the polymer increases, the  systems cross the threshold from unentangled to entangled dynamics. For entangled systems the dynamical rescaling approach that we propose is modified to include a one-loop perturbation that accounts for the presence of entanglements. Simulations of soft colloidal liquids are performed for entangled systems, and the rescaling applied to predict diffusion. 

\begin{table}[h]
\centering
\caption{Polyolefin melts UA MD simulation parameters}
\bigskip
\begin{tabular}{lcccc}\hline \hline
\mbox{{\it System}}      
& $T$[K] & $\rho$[sites/\AA$^{3}$] & $(R_g^{UA})^2$[\AA$^2$] & $(R_g^{FRC})^2$[\AA$^2$] \\
\hline
PE 30$^a$    &    400   &    0.0317094     &   63.5695  &    81.7544   \\ 
PE 44$^a$    &    400   &    0.0323951     &  110.3197  &   127.6856   \\ 
PE 48$^b$    &    450   &    0.0314487     &  111.0832  &   140.8119   \\ 
PE 66$^a$    &    448   &    0.0328993     &  177.5348  &   199.8812   \\ 
PE 78$^b$    &    450   &    0.0321465     &  205.9221  &   239.2607   \\ 
PE 96$^a$    &    448   &    0.0328194     &  281.7989  &   298.3301   \\ 
PE 122$^b$   &    450   &    0.0325479     &  346.2655  &   383.6526   \\ 
PE 142$^b$   &    450   &    0.0326600     &  420.7070  &   449.2852   \\ 
PE 174$^b$   &    450   &    0.0327680     &  525.1816  &   554.2975   \\ 
PE 224$^b$   &    450   &    0.0328835     &  690.5038  &   718.3791   \\ 
PE 270$^b$   &    450   &    0.0329520     &  856.4648  &   869.3342   \\ 
PE 320$^b$   &    450   &    0.0330034     &  980.1088  &  1033.4158   \\ 
\hline \hline
\multicolumn{4}{l}{$^a$ from Refs. \cite{MONDE,JARAM,HEINE}; $^b$ from Ref. \cite{Mavran,Mavran1}}
\end{tabular}
\label{TB:parameters_UAMDa}
\end{table}

Details about  our MS MD simulations have been reported in previous papers of ours and will not be repeated here.\cite{jaypaper,shortivan,multis} Briefly, MS MD simulations were implemented in the microcanonical $(N,V,E)$ ensemble on a cubic box with periodic boundary conditions. We used reduced units such that all the units of length were scaled by $R_g$ ($\tilde{r}=r/R_g$) and energies were scaled by $k_B T$. Temperature and radius-of-gyration were utilized for dimensionalizing  the results obtained from the MS MD simulations, after they were performed. 

The number of particles, i.e. polymer chains, in our simulations varies from $1728$ ($N=40$) to $85184$ ($N=1000$) depending on the system. This number is determined by the box size, which is larger than twice the range of the potential, and by the liquid density. The potential is long-ranged, due to the many-body effects entering through the OZ equation. 

Each simulation evolves for about $50,000$ computational steps. For the entangled melts the potential is longer ranged than for the unentangled systems, and therefore it is cut at larger distances requiring a bigger box size. The reduced density used in simulations, $\rho^{sim} = \rho R_g^3$ where $\rho$ is the site density,  varies around $1$ for unentangled melts, and exceeds $2$ for weakly entangled melts. A typical MS simulation takes between $2$ hours ($N=40$) to $4$ days ($N=200$) on one CPU workstation, while using the code that works in parallel, the computational time is further reduced.

\subsection{Interparticle potential}
\label{SX:ip}
The pair potential  acting between two effective coarse-grained units is formally derived from the colloidal representation of the liquid, specifically $h(r)$, using an hyper-netted-chain (HNC) closure approximation to the Ornstein-Zernike equation.\cite{SMPLQ} This approximation is known to work well for  liquids of particles interacting through a soft potential.\cite{mcquarrie} 
The potential input to the MS MD simulation, $v^{cc}(r)$, is derived from the total correlation function for the soft colloidal representation of the liquid, 
$h^{cc}(r)$, defined in the limit of long chains, $N\rightarrow \infty$, as in Eq.(\ref{EQ:2}).
The potential is calculated using the hypernetted chain approximation as
\begin{equation}
\beta v^{cc}(r) = h^{cc}(r)-ln[1+h^{cc}(r)]-c^{cc}(r) \ .
\label{EQ:HNC}
\end{equation}
Here the direct correlation function, $c^{cc}(r)$ is given in reciprocal space in terms of $h^{cc}(k)$ as:
\begin{equation}
c^{cc}(k) = \frac{h^{cc}(k)}{1+\rho_{ch} h^{cc}(k)} \ .
\label{EQ:ccck}
\end{equation}

It is important to define the correct potential acting between the coarse-grained units to achieve a realistic representation of the large scale properties of a system through MS MD. Because coarse-grained potentials result from the mapping of many-body interactions into pair interactions, through the averaging over microscopic degrees of freedom, they are parameter dependent. During the coarse-graining procedure, the potential acting between microscopic units, which is given by the Hamiltonian of the system, reduces to an effective potential, which is a free energy in the reference system of the microscopic coordinates. The coarse-grained potential so obtained contains contributions of entropic origin due to the microscopic, averaged-out degrees of freedom and is therefore state-dependent.  This can be observed in the form of the total correlation function between coarse-grained sites, Eq.(\ref{EQ:2}) from which the potential is derived. The correlation function explicitly includes the structural and thermodynamic parameters of the polymer, i.e. the radius-of-gyration, density screening length, and number density. The temperature enters directly through  Eq.(\ref{EQ:HNC}) and indirectly through the molecular parameters, such as $R_g$.

\subsection{Results from mesoscale molecular dynamics simulations}
\label{SX:rf}

\begin{figure}[]
\centering
\includegraphics[scale=0.35]{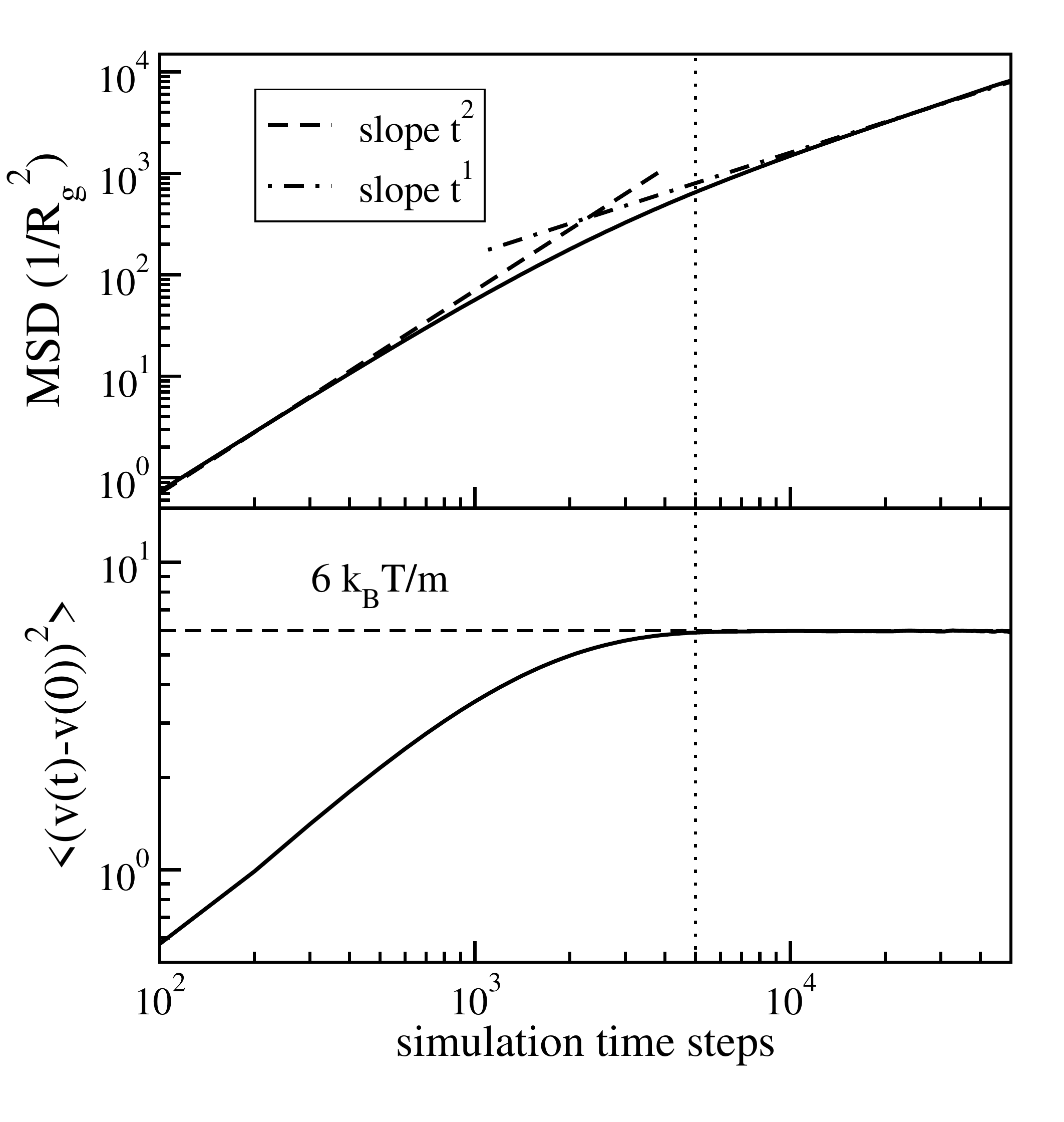}
\caption{Top panel: cm mean-square displacement (solid line) from MS MD simulation in reduced units as a function of simulation time steps for the $PE 44$ melt sample. The slopes for the ballistic and the diffusive regimes are shown as dashed and dot-dashed lines correspondingly. Bottom panel: cm velocity time correlation function showing when the inertial term becomes negligible. The asymptotic value of $6 \ k_BT/m$ is depicted as a dashed line.}
\label{FG:MS_MSD_vtcf}
\end{figure}

Before entering the details of applying our rescaling approach we focus on the "raw" dynamics obtained directly from the MS MD simulations. Fig.~\ref{FG:MS_MSD_vtcf} displays the mean-square displacement for the MS MD simulation of a polyethylene melt with $N=44$. At short times the inertial term in the Langevin equation is dominant as the particles undergo ballistic dynamics, while in the long-time regime the system crosses over to diffusive dynamics. The diffusion coefficient is higher than the value measured in UA MD simulations, and the transition from ballistic to diffusive regime happens after about $5000$ simulation steps (dot-dot vertical line on top panel of Fig.~\ref{FG:MS_MSD_vtcf}), which corresponds to a distance of roughly $30 R_g$. Such a large distance reflects the fact that in MS MD simulation the point particles interact through a very soft potential and the density is also very low. Because the particle has to "collide" many times to undergo the crossover to diffusive dynamics the latter takes place at a large lengthscale. 

Moreover, the bottom panel in Fig.~\ref{FG:MS_MSD_vtcf} displays the velocity correlation function $\left<(v(t)-v(0))^2\right>$ and shows that, consistent with the mean-square-displacement, once more the inertial term becomes negligible at the same crossover time that the diffusive regime sets in. Since our MS MD simulation are performed at equilibrium, the avarage kinetic energy per particle $\left<m v^2(t)/2\right> = 3/2\, k_BT$, and therefore
\begin{eqnarray}
\lim_{t\rightarrow\infty} \left<(v(t)-v(0))^2\right> &=& \lim_{t\rightarrow\infty} \left[2\left<v^2(t)\right>-2\left<v(t) v(0)\right>\right] \nonumber\\
&=& 6 k_BT/m\, .
\end{eqnarray}

The dynamical transition is displayed as a dashed line on the bottom panel of Fig.~\ref{FG:MS_MSD_vtcf}, taking into account that our simulations are in reduced units and $m=1$, $k_BT=1$. The figure shows that the velocity autocorrelation function reaches it's asymptotic value at about the same time as the diffusive regime sets in for the mean-square-displacement.

\section{Application of the Rescaling Procedure}
\label{SX:ao}
As stated above, the accelerated dynamics that is a consequence of the coarse graining of the system can be rescaled by taking into account the two main effects of the procedure, namely the change in the entropic contribution to the free energy of the system due to the averaging of the internal degrees of freedom and  the change in the friction coefficient due to the different shapes of the molecule in the two different levels of coarse-graining. The difference in shape relates to the change in the molecular surface available to the surrounding molecules, and to the correlation of the random forces generated 
by intermolecular collisions.

The first rescaling is given by the inclusion \textit{a posteriori} of the internal degrees of freedom, averaged out during the coarse-graining procedure, as a correction term in the free energy of the system, which accounts for the difference in entropy. The energy correction affects the time of the measured dynamics as the change from a bead-spring description to a soft-colloid representation leads to the rescaling of the  time reported in Eq.(\ref{EQ:4}), also taking into account the fact that because the potential is expressed in normalized quantities, the simulation runs using reduced units of energy, $k_BT=1$ and the normalized length $r/R_g$. 

The second rescaling  of the dynamics is calculated starting from the ratio between the friction coefficients in the two coarse-grained representations, as described in Eq.(\ref{dmsris}).

\subsection{Calculations of the monomer friction coefficient,~$\zeta_m$}
\label{pip}
Because our formalism maps the Lennard-Jones liquid described by the UA MD simulation into a liquid of polymers interacting through a monomer hard-core repulsive potential, it is necessary to define an effective hard-core diameter, $d$. This is done by requiring the friction of the chain with $N=44$ to follow the expected scaling behavior for the diffusion of an unentangled polymer chain, $D=k_B T/ (N \zeta_m)$. Since all except two of the atoms in our PE chains are $CH_2$ monomers, we assume that the potential is identical for all the units along the homopolymer chain. Moreover we assume that the range of the repulsive interaction, $d$, is independent of liquid density.\cite{hardsphere}

\begin{figure}[]
\centering
\includegraphics[scale=0.30]{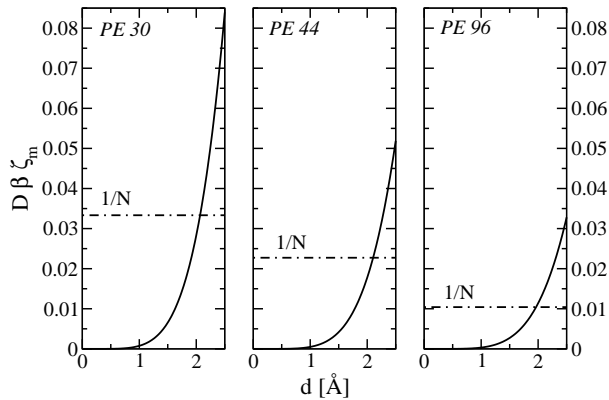}
\caption[Plot of $\zeta(d)$]{Plot of $D \beta \zeta_m$ as a function of hard core diameter $d$. From left to right, the three panels show curves for polyethylene melt with $N=30$, $N=44$, and $N=96$.}
\label{FG:zeta_of_d}
\end{figure}

Among the different samples, we selected the chain with $N=44$ to optimize $d$, because this sample follows unentangled dynamics while the polymer is long enough to obey the Gaussian intramolecular distribution of monomer positions, which justifies the analytical form of the intramolecular structure factors used in our formalism.
Fig.~\ref{FG:zeta_of_d} displays the monomer friction coefficient, from Eq.(\ref{ciccio}), expressed as the dimensionless quantity $D \beta \zeta_m$, as a function of hard sphere diameter $d$, for polyethylene melts of three different degrees of polymerization. The  $1/N$ scaling is reported as a dot-dashed line in the figure.

In these calculations, the numerical values of $N$, $\rho$ and $R_g$ were taken from the data of the UA MD simulation against which the proposed approach is tested. The value of the radial distribution function at the contact was set to $g(d)=1/2$, which is the conventional value assumed in the PRISM thread theory for polyethylene chains. This value is intermediate between zero and the first solvation shell value. The optimized hard-core diameter for $N=44$ is $d=2.1\AA$, which is  an intermediate value between the bond length, $l=1.54$   \AA, and the Lennard-Jones $\sigma$-parameter, $\sigma=3.95$ \AA, in the intermonomer potential of the UA MD.\cite{JARAM,HEINE} The unentangled scaling is fulfilled for PE30 at $d=2.07$ \AA \ and for PE96 at $d=1.96$ \AA, which are close to the one for PE44. Because the PE96 sample has a degree of polymerization that is close to the entanglement value of $N_e=130$, its dynamics is likely to be in the crossover regime where the effect of entanglements start to be felt, since the transition from the unentagled to the entangled dynamics is very broad. 

Table \ref{TB:mon_hard_sphere} displays the numerical values of the dimensionless monomer friction coefficient, $D \beta \zeta_m$, for polymeric liquids with different degree of polymerization, $N$, across the untentengled-to-entangled transition. For both unentangled and entangled systems the hard sphere diameter has been fixed to the value of the unentangled ones, $d=2.1$\AA, so that the intermolecular monomer potential is not changed as a function of $N$. While for unentangled systems the monomer friction coefficient was calculated from Eq.(\ref{ciccio}), for entangled    chains we adopted a perturbative approach to account for the effect of entanglements.
Let's denote
\begin{equation}
D\beta \zeta_m = J(\rho, N, R_g, d) \, ,
\label{EQ:bulky2}
\end{equation}
where for unentangled systems $J(\rho, N, R_g, d) \approx N^{-1}$. Following a one-loop perturbation, and including the definition of the diffusion coefficient for a macromolecule comprised of $N$ monomers with $\zeta_m$ the monomer friction coefficient  $D=(\beta \zeta_m N)^{-1}$, the normalized and perturbed friction coefficient becomes
\begin{equation}
D\beta\zeta_m' = N D\beta \zeta_m J(\rho, N, R_g, d) = N (D\beta \zeta_m)^2\ .
\label{entangly}
\end{equation}

The one loop perturbation is in the spirit of the reptation model where both the chain reptating and the chains involved in the entanglements  relax with the same diffusive mechanism: each brings a $N^{-1}$ scaling contribution, which is the trademark of polymer Brownian motion. Interestingly, in our model the diffusion coefficient of entangled polymers under certain fixed monomer density and temperature shows apparent scaling exponents  different from  ${- 2}$ (see Fig.~\ref{FG:diffusion1}). The resulting scaling exponents emerge cumulatively from output of mesoscale simulations and both steps of rescaling.

Because Eq.(\ref{entangly}) applies only when the systems are entangled, to predict the diffusive behavior of new samples it is necessary to estimate a priori the crossover degree of polymerization, $N_e$. Several methods have been presented in the literature to estimate $N_e$ from thermodynamic conditions and molecular parameters.\cite{ennee} Those methods provide similar values of $N_e$. Moreover, the expressions for the unentangled and entangled frictions, Eqs.(\ref{ciccio}) and (\ref{entangly}), predict values that differ only slightly in the crossover region, as it is shown in Fig.~\ref{FG:diffusion1} in this paper. In this way, selecting the unentangled expression to represent entangled systems, or viceversa, in the crossover region would result in small inconsistencies in the calculated diffusion coefficients. 

Table \ref{TB:mon_hard_sphere}  includes intra- ($D\beta\zeta_m(\omega_{\alpha \gamma})$) and inter-molecular ($D\beta\zeta_m(h_{\alpha \gamma})$) contributions to the monomer friction coefficient, as well as the self intramolecular contribution, $D\beta\zeta_m(\omega_{\alpha\alpha})$. In general, the calculated total friction is comparable in magnitude to the self intramolecular contribution, $D\beta\zeta_m(\omega_{\alpha\alpha})$. Moreover, the total intramolecular contribution, $D\beta\zeta_m(\omega_{\alpha,\gamma})$, is of the same order of magnitude of the itermolecular contribution, $D\beta\zeta_m(h_{\alpha,\gamma})$, but with the opposite sign, which is reasonable as the liquid is almost incompressible. This result shows that the conventional approximation of replacing the structure factor, $S_{\alpha,\gamma}(k)$, by the single chain analog, $\omega_{\alpha,\gamma}(k)$, can lead to errors in the evaluation of the memory function for macromolecular liquids.\cite{Hess} The table also displays the value of the dimensionless friction coefficient $N D \beta \zeta_m$, which for unentangled systems should be $\approx 1$. As expected, we see deviation from the unentangled behavior in the very short chains and in the crossover to entangled dynamics at $N \approx 100$.

\begin{table}[h]
\centering
\caption{Monomer friction coefficient contributions with hard sphere potential for $d = 2.1$\AA} 
\bigskip
\begin{tabular}{lccccc}\hline \hline
\mbox{{\it System}}      
& $D\beta\zeta_{m}(\omega_{\alpha \alpha}) $ & $D\beta\zeta_{m}(\omega_{\alpha \gamma}) $  
& $D\beta \zeta_{m}(h_{\alpha \gamma})$ & $D\beta\zeta_{m}$ & $N D\beta\zeta_{m}$\\
\hline
PE 30   &  0.03378    & 0.08484    &  -0.04883  &   0.03601  &    1.0804    \\
PE 44   &  0.02744    & 0.06244    &  -0.03991  &   0.02253  &    1         \\
PE 48   &  0.03101    & 0.07984    &  -0.04904  &   0.03080  &    1.4782    \\
PE 66   &  0.02539    & 0.05703    &  -0.03823  &   0.01880  &    1.2409    \\
PE 78   &  0.02632    & 0.06287    &  -0.04179  &   0.02107  &    1.6439    \\
PE 96   &  0.02239    & 0.04766    &  -0.03325  &   0.01441  &    1.3834    \\
PE 122  &  0.02400    & 0.05533    &  -0.03831  &   0.01701  &    2.0757    \\
PE 142  &  0.02245    & 0.04967    &  -0.03501  &   0.01466  &    2.0815    \\
PE 174  &  0.02193    & 0.04827    &  -0.03438  &   0.01389  &    2.4170    \\
PE 224  &  0.02133    & 0.04664    &  -0.03359  &   0.01304  &    2.9219    \\
PE 270  &  0.02039    & 0.04344    &  -0.03164  &   0.01180  &    3.1849    \\
PE 320  &  0.02184    & 0.04956    &  -0.03585  &   0.01371  &    4.3874    \\
\hline \hline
\end{tabular}
\label{TB:mon_hard_sphere}
\end{table}

\subsection{Calculation of the friction coefficient of a soft colloid, $\zeta_{soft}$}
Starting from Eq.(\ref{EQ:zeta_soft_mean_force}) we calculated the friction coefficient, $D \beta \zeta_{soft}$, for polymer liquids represented as soft colloidal particles. Table \ref{TB:Soft_MFP} shows the dimensionless friction coefficient for several systems. The molecular parameters,  $N$ and $R_g$, and the thermodynamic conditions of density $\rho$ and temperature $T$, are taken from the UA MD simulations, see Table \ref{TB:parameters_UAMDa}.
\begin{table}[h]
\centering
\caption{Soft colloids friction coefficient contributions}
\bigskip
\begin{tabular*}{230pt}{@{\extracolsep{\fill}}lccc}\hline \hline
\mbox{{\it System}}      
& $D\beta\zeta_{\text{soft}}^{\text{self}} $  & $D\beta\zeta_{\text{soft}}(h^{cc})$ & $D\beta\zeta_{\text{soft}}(1 + h^{cc})$\\
\hline
PE 30    &  0.044273     &  -0.029932    &   0.014341      \\
PE 44    &  0.020769     &  -0.012441    &   0.008328      \\
PE 48    &  0.024639     &  -0.015289    &   0.009349      \\
PE 66    &  0.012619     &  -0.006659    &   0.005960      \\
PE 78    &  0.012019     &  -0.006254    &   0.005765      \\
PE 96    &  0.008055     &  -0.003712    &   0.004344      \\
PE 122   &  0.007423     &  -0.003334    &   0.004089    \\
PE 142   &  0.006159     &  -0.002612    &   0.003546    \\
PE 174   &  0.005218     &  -0.002108    &   0.003109    \\
PE 224   &  0.004298     &  -0.001648    &   0.002650    \\
PE 270   &  0.003660     &  -0.001351    &   0.002310    \\
PE 320   &  0.003521     &  -0.001288    &   0.002233    \\
\hline \hline
\end{tabular*} \\
\label{TB:Soft_MFP}
\end{table}
The dimentionless friction coefficient for these systems is $D\beta\zeta_{\text{soft}}(1 + h^{cc})\approx0.002-0.01$, while we would expect $D\beta\zeta_{\text{soft}}(1 + h^{cc})\approx 1$ for unentangled systems, see Table \ref{TB:Soft_MFP}. These data show that the theoretically calculated friction coefficient (without rescaling) for the soft colloidal systems greatly underestimate the friction coefficient, as also observed in the MS MD simulations, and hence give rise to accelerated dynamics as discussed previously. It also shows that intra- and inter-molecular contributions to the friction coefficient are comparable in magnitude: both of them need to be taken into account when calculating dynamical properties of polymer melts.

\subsection{Results from the rescaling procedure. Comparison with simulation and experimental data}
\label{SX:rft}
Some of the results reported in this section were already briefly presented in our short paper.\cite{shortivan} Our discussion here makes use of some of those data as a starting point to illustrate with an example the details of the proposed rescaling procedure and highlight its strengths and weaknesses.

In order to rescale the unrealistic fast dynamics of MS MD simulations  we applied our rescaling procedure and compare the predicted dynamics with data from UA MD simulations and experiments. We use as input parameters the thermodynamic conditions and molecular parameters of each sample under study. The rescaling procedure is given by Eq.(\ref{dmsris}), where $D^{MS}$ is the diffusion coefficient from the MS-MD simulation, the soft colloid friction coefficient is calculated using Eq.(\ref{EQ:zeta_soft_mean_force}), and the monomer friction coefficient is given by Eq.(\ref{ciccio}) for unentangled chains, and Eq.(\ref{entangly})with Eq.(\ref{ciccio}) for entangled ones.  Eq.(\ref{dmsris}) depends on the temperature and density of the system investigated and on its molecular radius-of-gyration.
 
Indirectly those parameter enter our procedure through the diffusion coefficient from mesoscale simulations, $D^{MS}$. Specifically temperature enters through the rescaling of the time, as the time step in the mesoscale simulation is adimensional and becomes dimensional once it is rescaled by the energy, following a well-established procedure. Moreover, thermodynamic and molecular parameters enter indirectly through the soft potential, Eq.(\ref{EQ:HNC}), which is parametric and includes density, temperature, and the molecular radius-of-gyration.

Finally, thermodynamic parameters enter directly through the definitions of the friction coefficients in the monomer and soft-sphere descriptions, Eqs.(\ref{ciccio}) and (\ref{EQ:zeta_soft_mean_force}) respectively. Specifically, the monomer friction coefficient is a function of   $\rho$, $N$, $R_g$, plus a hard sphere diameter, $d$, which is used to map the Lennard-Jones potential of the united atom simulation onto a repulsive hard-core potential with an effective bead diameter. The  hard sphere diameter $d$ is assumed to be independent of the thermodynamic conditions, for the range of temperature and density simulated here, and constant for all the monomers in the homopolymer chain. The criteria of choosing numerical value for $d$ have been already explained and discussed.

In an analogous way, the soft-sphere friction coefficient depends on the chain number density, which relates to the monomer number density through $N$ as $\rho_{ch} = \rho/N$, and $R_g$ is the radius of gyration of the polymer chain. It also depends on the density fluctuation length scale $\xi_\rho$, which is expressed as a function of $R_g$ and $\rho_{ch}$ as $\xi_\rho = {R_g}/(\sqrt{2}+2\pi\rho_{ch}R_g^3)$, and on the parameter $h_0=h(k=0)=-(1-2\xi_\rho^2/R_g^2)/\rho_{ch}$. In fact, the dimensionless combination $D\beta\zeta_{\text{soft}}$ is determined by only three parameters: $\rho$, $N$ and $R_g$. In conclusion, once thermodynamic parameters, $R_g$ and $d$ are defined, there are no adjustable parameters in our method.

The predicted diffusion coefficients from our rescaled MS MD simulations are  in good agreement with the data for all the test systems.
As an example, Table \ref{TB:D_MS_MD} displays the diffusion coefficients obtained directly from the MS MD, $D^{MS}$, once they are rescaled to include the internal degrees of freedom, and after the second rescaling of the friction, $D_{cm}$, as well as the values of the diffusion coefficient from the UA MD simulations, $D^{UA}$, against which we compare our predicted diffusion.  The table shows that while the initial values of the diffusion are orders of magnitude larger than the data  from UA MDs, the rescaled coefficients are very close to the real values. For the entangled systems  we adopt the perturbative approach described in Section \ref{pip} obtaining predicted values that are in quantitative agreement with the UA MD simulations. Entangled samples are from references \cite{Mavran,Mavran1}, and are mostly in the weakly entangled regime. For these samples the UA MD simulations include a small number of chains: $n=40$  for $N=78$, $n=22$ for $N=142$, $n=32$ for $N=174$, $N=224$, $N=270$ and $N=320$, with $n$ the number of chains in the simulation and $N$ the degree of polymerization. These numbers show one advantage of adopting a coarse-grained description as typically our samples include thousands of chains. Simulating a large ensemble of molecules is necessary, for example, when the goal is to investigate large-scale fluctuations or the relative relevance of intra- vs inter-molecular contributions to the dynamics. 

\begin{table}[h]
\centering
\caption{Diffusion coefficients in \AA$^2$/ns from MS MD compared with UA MD simulation}
\bigskip
\begin{tabular*}{230pt}{@{\extracolsep{\fill}}lcccc}\hline \hline
\mbox{{\it System}}      
& $T$[K]& $D^{MS}$   & $D_{cm}$   & $D^{UA}$ \\
\hline
PE 30 &   400   &  4.44$\times 10^3$  &    58.9   &        82.9    \\
PE 44 &   400   &  5.29$\times 10^3$  &    44.5   &        46.0    \\
PE 48 &   450   &  5.80$\times 10^3$  &    36.7   &        50.8    \\
PE 66 &   448   &  6.04$\times 10^3$  &    29.0   &        31.8    \\
PE 78 &   450   &  6.73$\times 10^3$  &    23.6   &        26.0    \\
PE 96 &   448   &  6.98$\times 10^3$  &    21.9   &        23.3    \\
PE 142&   450   &  8.45$\times 10^3$  &    6.92   &        7.93    \\
PE 174&   450   &  8.51$\times 10^3$  &    4.53   &        5.70    \\
PE 224&   450   &  8.80$\times 10^3$  &    2.73   &        3.28    \\
PE 270&   450   &  9.39$\times 10^3$  &    2.14   &        2.06    \\
PE 320&   450   &  8.73$\times 10^3$  &    1.03   &        1.30    \\
\hline \hline
\end{tabular*}
\label{TB:D_MS_MD}
\end{table}

\begin{figure}
\centering
\includegraphics[scale=0.4]{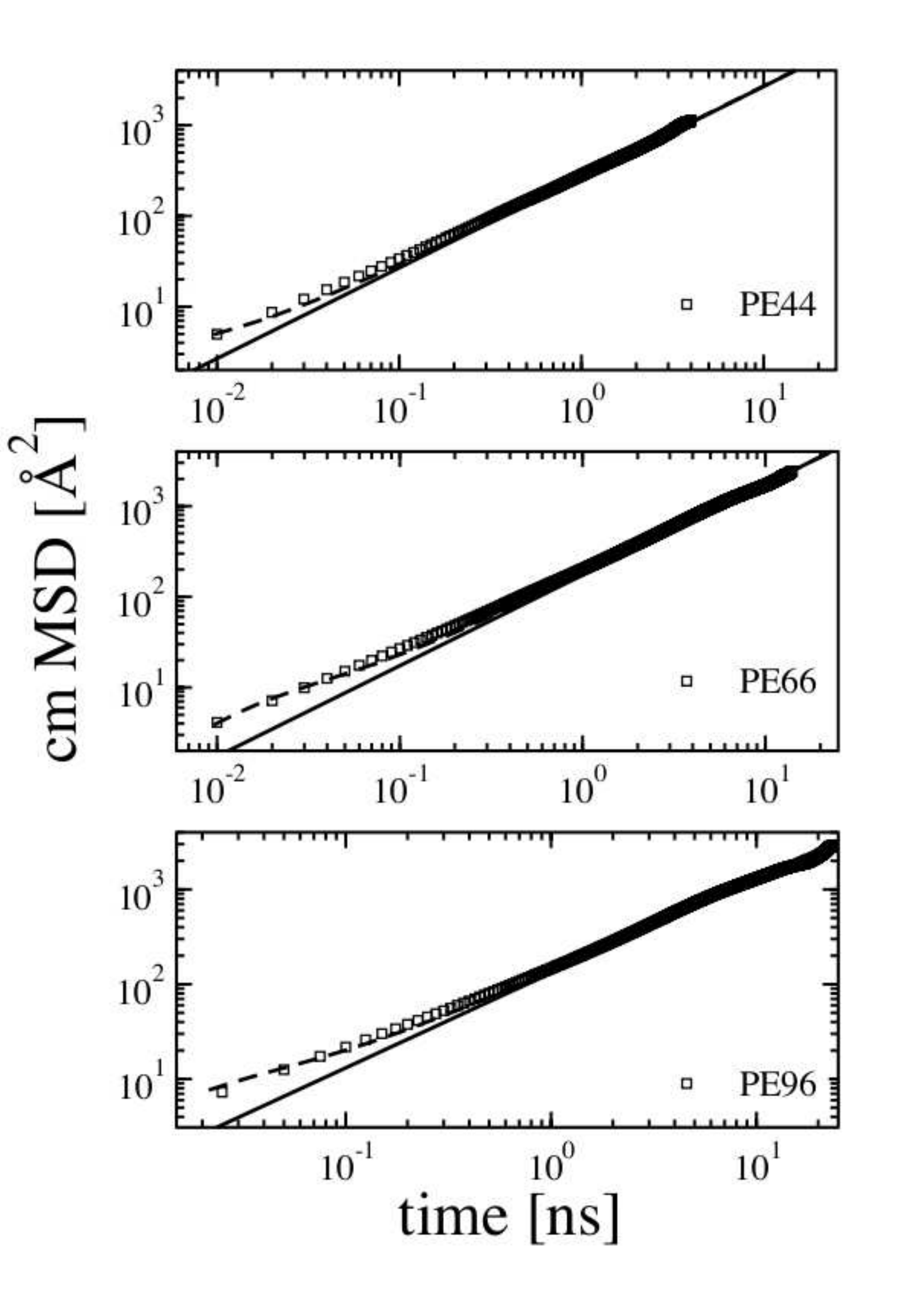}
\caption[]{ Plot of mean-square displacement as a function of time for unentangled $PE$ melts. The rescaled MS MD simulation (line) is compared with UA MD simulation (symbols) for $N=44, 66, 96$. Also shown is the outcome of the theory for cooperative dynamics (dashed lines).}
\label{FG:MSD_PE_approximations}
\end{figure}

Fig.~\ref{FG:MSD_PE_approximations} illustrates how our approach can be used to calculate dynamics also in the short time regime. The figure shows the mean-square-displacement of the center-of-mass from UA MD in comparison with the one calculated from the diffusion coefficient rescaled from the MS MD. The agreement is quantitative in the long time regime. In the short time regime, the UA MD simulation data exhibit a subdiffusive behavior, even if polymers are unentangled. In a series of papers we have shown that the subdiffusive regime is a consequence of  the presence of cooperative  dynamics involving several polymer chains moving in a correlated way  inside the dynamically heterogeneous liquid of macromolecules.\cite{manychains,Marina} A detailed  discussion of this phenomenon, which is of intermolecular origin, has been provided before and will not be  repeated here. Our theory, the Cooperative Dynamics Generalized Langevin Equation (CD-GLE), needs as an input the diffusion coefficient and predicts the subdiffusive behavior for times shorter than the longest correlation time as a function of the number, $n' \propto \sqrt{N}$, of macromolecular chains moving in a cooperative way. Fig.~\ref{FG:MSD_PE_approximations} shows the results from our CD-GLE calculations as dashed lines. Here, the input monomer friction  coefficient is calculated from MS MD using the rescaling procedure. The number of chains undergoing cooperative dynamics is $n'=30$ for $N=96$, $n'=25$ for $N=66$, and $n'=14$ for $N=44$.

The subdiffusive behavior shown in UA-MD data is not visible in the rescaled MS MD data as the dynamics are accelerated. The effective temperature experienced by the polymer is much higher than the temperature in UA MD simulations as the energy is not dissipated in the internal degrees of freedom.

Finally we discuss the calculations of the monomer and soft-colloid friction coefficients for a set of polyethylene chains investigated experimentally.\cite{Richter,maybeothers,maybeothers1,maybeothers2,maybeothers3} The experimental data do not report the values of $R_g$ at the desired thermodynamic conditions, $T=509$ K and density $\rho=0.0315302$ sites/\AA$^3$, while it is known that the chain conformation, and $R_g$, are temperature dependent. To calculate the input parameters for our MS MD simulations we adopt a freely rotating chain model, for which the mean-square end-to-end polymer distance is given by\cite{Yamakawa}

\begin{eqnarray}
\left<R_{ete}^2 \right>=N l^2\left[ \frac{1+g}{1-g}-\frac{2g}{N}\frac{1-g^N}{(1-g)^2}  \right] \ ,
\end{eqnarray}
and $R_g^2\approx \left<R_{ete}^2 \right>/6$ for a chain with Gaussian statistics. For polyethylene melts  at this temperature the stiffness parameter is $g=0.785$. \cite{Richter}
For the samples investigated in the UA MD simulations, a comparison of the theoretical values of $R_g$ using the freely-rotating-chain model (FRC) and the values measured directly from the UA MD, which are  both reported in Table \ref{TB:parameters_UAMDa}, show a reasonable agreement. The agreement is particularly good for the samples that have long chains because for them the hypothesis of a Gaussian intramolecular distribution is well justified. In this framework, the FRC model provides a reliable description of the chain intramolecular structure.

The values of the monomer and soft-colloid friction coefficients for the experimental samples, calculated from Eqs. (\ref{ciccio}) and (\ref{EQ:zeta_soft_mean_force}) respectively, are presented in Table \ref{TB:zeta_EXP}. The Table shows the large difference between the predicted dimensionless friction coefficients, $D \beta \zeta_{soft}$ and $N D \beta \zeta_m$, for the same macromolecule coarse-grained at two different length scales. From the values displayed in Table \ref{TB:zeta_EXP} we calculate the rescaling factor for the friction coefficient measured in MS MD simulations, following the procedure described in this paper. Because the data have different thermodynamic parameters of density and temperature, their scaling behavior cannot be inferred from their plot, even if an apparent $N^{-1}$ scaling is followed by the unentangled samples and the typical reptation $N^{-2}$ scaling  by the entangled ones.

\begin{table}[h]
\centering
\caption{Theoretically calculated dimensionless friction coefficient for monomer ($d=2.1$\AA) and soft colloid with $R_g^{FRC}$ for experimental samples}
\bigskip
\begin{tabular}{lccc}\hline \hline
\mbox{{\it System}}      
& $(R_g^{FRC})^2$[\AA$^2$] & $D\beta \zeta_{soft} $ & $N D\beta \zeta_m$ \\
\hline
PE 36    &  101.4350     &    0.007846    &    0.5153        \\
PE 72    &  219.5710     &    0.004927    &    0.8946        \\
PE 106   &  331.1465     &    0.004543    &    1.8407       \\
PE 130   &  409.9056     &    0.003497    &    1.5354    \\
PE 143   &  452.5669     &    0.003318    &    1.6822     \\
PE 192   &  613.3669     &    0.002828    &    2.2412     \\
PE 242   &  777.4485     &    0.002500    &    2.8186     \\
\hline \hline
\multicolumn{3}{l}{$T=509$K, $\rho=0.0315302$ [sites/\AA$^3$]}\\
\end{tabular}
\label{TB:zeta_EXP}
\end{table}

\begin{table}[h]
\centering
\caption{Predicted diffusion coefficients in \AA$^2$/ns from MS MD and experimental data from
 Refs. \cite{Richter,maybeothers,maybeothers1,maybeothers2,maybeothers3}}
\bigskip
\begin{tabular}{lcccccccc}\hline \hline
{\it System} & PE36 & PE72 & PE106 & PE130   & PE143   & PE192   & PE242  \\
$D_{cm}$ & 111 & 50 & 17 & 14 & 11 & 5.6 & 3.2 \\
$D^{exp}$ & 120 & 41 & 14 & 12 & 8.6 & 6.5 & 4.5 \\
\hline \hline
\end{tabular}
\label{TB:parameters_EXP}
\end{table}


\subsection{Theoretical predictions of diffusion coefficients for polyethylene samples}
\begin{figure}
\centering
\includegraphics[scale=0.28]{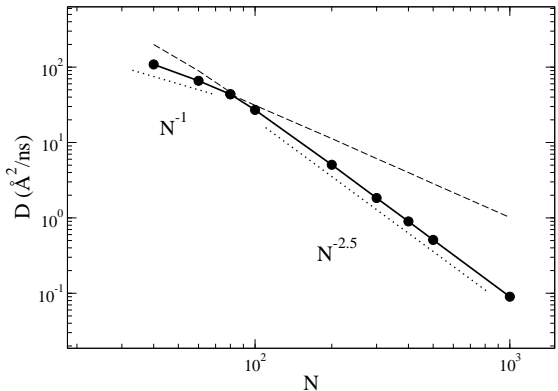}
\caption[]{Plot of diffusion coefficients (symbols) as a function of degree of polymerization, $N$.  Also shown are the scaling exponents for our unentangled, $N^{-1}$ (dotted line), and entangled systems, $N^{-2.5}$ (dotted line).}
\label{FG:diffusion1}
\end{figure}

In this section we report theoretical predictions from rescaled mesoscale simulations of the diffusion coefficients for a series of PE samples for which data of chain dynamics, either from simulations or from experiments, are not available in the literature. The degree of polymerization of each sample is not larger than the ones already investigated. However, because there are no data to fit any parameter, these calculations illustrate the predictive power of our method. Diffusion coefficients calculated by combining the mesoscale simulations with the rescaling procedure presented in this manuscript, are displayed in Fig.~\ref{FG:diffusion1} as a function of the degree of polymerization. 

The set of MS MD simulations is performed for $N=40, 60, 80, 100, 200, 300, 400, 500$ and $1,000$, at constant monomer density $\rho_m = 0.0329497$ [sites/\AA$^3$] and temperature $T=450 \ K$ for all samples. Values of $R_g$ are calculated using a freely rotating chain model. The hard sphere diameter is fixed to the value reported in the previous sections for PE, $d=2.1$\AA, and the pair distribution function at contact is $g(0)=1/2$ as described early on in this paper. While the simulations of the small samples can be performed on a single CPU machine, for systems with a higher degree of polymerization is convenient to adopt parallel computing. For those systems, simulations were run using the LAMMPS code\cite{LAMMPS}, with our potential as an input, remotely on a 64 CPU machine available through the TeraGrid\cite{TERAGRID}. For the PE1000 sample, which included 85,184 molecules, results were obtained after one week of calculations. By comparison with our single CPU calculations, running the simulation in parallel reduces the computer time by a factor of $10^2$. The number of particles in the simulation is determined by the length of the box size, which for PE1000 sample is equal to $24$ $R_g$, i.e. larger than twice the range of the potential,  to eliminate molecular self-interaction through the periodic boundary conditions.

While we assume that small changes of density and temperature do not affect the hard-core diameter, $d$, even a small difference in $\rho_m$ can noticeably change the prefactor in Eq.(\ref{dmsris}). The monomer friction coefficient is calculated using Eq.(\ref{ciccio}) for unentangled systems and Eq.(\ref{entangly}) for entangled ones. The full lines in Fig.~\ref{FG:diffusion1} show the equations used in the calculation, while the dashed lines represents the prediction of the equation for the entangled system in the unentangled region, and the prediction of the unentangled equation in the entangled region. In the crossover regime, $N\approx 100$ for PE, both expressions lead to very similar results, providing a smooth crossover between the two equations.

The predicted values of the diffusion coefficient appear to be consistent with the known experimental behavior. The diffusion coefficients of unentangled chains ($N < 130$) follow the scaling behavior of the Rouse approach, while the entangled chains show a scaling with degree of polymerization of  $-2.5$.  Although the latter scaling exponent disagrees with the ``reptation model", it is known that experimental samples of weakly entangled chains also show a scaling exponent of $-2.5$. For those polymer chains, which are just across the transition from unentangled to entangled dynamics, constraint release and "tube" fluctuations are relevant. The observed scaling behavior is also consistent with the scaling of the viscosity observed experimentally.\cite{Lodge} The advantage of our method with respect to UA-MD simulations is that even in the case of long entangled chains it is possible to include a large number of molecules, improving the statistics of calculated correlation functions. Overall this plot shows that it is possible to provide reasonable predictions of large scale dynamical properties by properly rescaling mesoscale simulations.

\section{Discussion and Conclusions}
\label{SX:da}
The need for developing a fundamental approach  to rescale dynamical data obtained from MS simulations of coarse-grained systems has been a long-standing problem from the time that  coarse-graining approaches started being developed. Because MS simulations are less computational demanding than atomistic simulations, it is possible to  investigate larger systems for longer times than in all-atom simulations, allowing one to extend the maximum time and length scales accessible through simulations and to improve the statistics of measured averaged quantities. Considering that the number of particles in a simulation should be large enough to ensure proximity to the thermodynamic limit, MS simulations of coarse-grained systems could become an indispensable tool to investigate the structure and dynamics of macromolecular liquids.

One advantage of a MS simulation of a coarse-grained system is that the simulation speeds up because of the averaging of the internal degrees of freedom, leading to a softer potential and allowing the study of longer timescales than in a fully atomistic simulation. This implies, however, that the dynamical properties resulting from the MS MD are faster than their real counterpart, for example the ones from UA MD and need to be rescaled.  

It is the common procedure to rescale the measured dynamics numerically by bringing a time correlation function to agree with the one measured in atomic level simulations, however we adopt a different strategy.  We have proposed a first principle approach to derive an analytical form of the rescaling procedure to be applied to the dynamics measured directly from MS MD of a coarse-grained polymer liquid. Our approach allows for the reliable prediction of the long-time diffusion of a polymer melt as it would be measured in an atomistic or UA-MD simulation. The rescaling procedure has been tested so far against simulations and experiments of polyethylene liquids both unentangled and entangled. Calculated diffusion coefficients for samples for which we do not have data either experimental or simulated, show consistent behavior.

We start by running MS MD simulations of coarse-grained polyethylene melts where each polymer is represented as a point particle. The analytical intermolecular potential, input to the MS MD,  is derived from the  Ornstein-Zernike equation with the hypernetted closure approximation. The correction term to the measured dynamics of the MS MD simulations, is calculated from the solution of the Generalized Langevin Equations written for the coarse-grained and for the monomer-level representations of the macromolecular liquid. Those equations are formally derived from the Liouville equation by assuming two different lengthscales characterizing the relevant slow dynamics, i.e. monomer and center-of-mass, onto which the Liouville equation is projected.

While the Mori-Zwanzig projection operator technique suggests a reliable criteria to select the proper length scale of coarse-graining for dynamical properties, the GLEs thus generated allow one to derive analytical forms of the rescaling contributions associated with the coarse-grained dynamical equations. The rescaling procedure includes two contributions, given by the changes in entropy and in the friction coefficient during coarse-graining. The entropic contribution emerges from the averaging of the internal degrees of freedom, while the friction is due to the change in shape, and as a consequence the change of the molecular surface exposed to the surrounding molecules. Both corrections depend on the thermodynamic conditions of the system simulated,  and on the molecular structure through the radius-of-gyration of the macromolecule. Thermodynamic and molecular quantities enter both directly through the rescaling equations and indirectly through the effective potential in the mesoscale simulations. In this way the dynamics predicted from the rescaling of each mesoscale simulation is specific of the system under study.

A feature of the coarse-graining models we study is the mapping of the polymeric liquid onto simple representations, which are isotropic. At the molecular level the polymer is described as a soft isotropic sphere. At the monomer level, the bead-and spring description affords equivalent beads in the chain, which is a reliable approximation due to the high number of statistically equivalent structural configuration of the molecule. Chain end effects enter in the model through the finite size of the polymer in the matrix representation of the equations. Moreover, because the monomers in a homopolymer are structurally identical, with the exception of the two end monomers,  the intermolecular monomer-monomer hard-core interaction potential is assumed to be identical for any pair of monomers, and each monomer is supposed to have identical friction coefficient.

Although the theoretical picture is straightforward, our approach has the advantage of being described in closed-form expressions, even if approximated, which allows for an analytical solution of the rescaling formalism. This has the potential of being useful in improving our understanding of the nature of coarse-graining procedures.

\section*{APPENDIX I: INTERNAL ENERGY CALCULATION FOR A FREELY-ROTATING-CHAIN MODEL}
\label{Appendix1}
The effective mean-force potential for one homopolymer composed of $N$ monomers can be expressed through the structural matrix $\mathbf{A}$ as
\begin{eqnarray}
U = \frac{3k_B T}{2 l^2} \sum_{i,j} \mathbf{A}_{i,j}\, \mathbf{r}_i \cdot \mathbf{r}_j& = &\frac{3k_B T}{2 l^2} \sum_{x,y,z} \sum_{i,j} x_i \mathbf{A}_{i,j} x_j \nonumber \\
&=& \frac{3 k_B T}{2\, l^2} \mathbf{r}^T \mathbf{A} \mathbf{r} \, ,
\end{eqnarray}
with the matrix $\mathbf{A}$ being real and symmetric, and diagonalized by the orthonormal matrix of the eigenvectors $\mathbf{Q}^{-1}=\mathbf{Q}^T$, so that
\begin{equation}
\mathbf{r}^T \mathbf{A} \mathbf{r} = \mathbf{\xi}^T \mathbf{Q}^{-1} \mathbf{A} \mathbf{Q} \mathbf{\xi} = \mathbf{\xi}^T \mathbf{\Lambda} \mathbf{\xi} \, ,
\end{equation}
where $\mathbf{\Lambda}$ is the matrix of the eigenvalues, and $\xi$ is the matrix of the normal modes defined by $\mathbf{r}=\mathbf{Q} \mathbf{\xi}$. 

In this model, the equilibrium distribution function is 
\begin{eqnarray}
\Psi_{eq}(\textbf{r}) &=& \textit{N}_x e^{-\frac{3}{2 l^2} x^T A x} \textit{N}_y e^{-\frac{3}{2 l^2} y^T A y} \textit{N}_z e^{-\frac{3}{2 l^2} z^T A z} \nonumber \\
&=& \textit{N} e^{-\frac{3}{2\, l^2} \mathbf{r}^T \mathbf{A}\, \mathbf{r}} =  \textit{N} e^{-\frac{1}{2} \mathbf{r}^T \mathbf{A'}\, \mathbf{r}} \, ,
\label{EQ:distribution}
\end{eqnarray}
where for convenience of notation we introduced the matrix $\mathbf{A'}=3 \mathbf{A}/l^2$.
Here $\textit{N}_x$ is the normalization factor, defined by enforcing $\int dx \Psi_x =1$, as 
\begin{equation}
N_x = \left(\frac{3}{2 \pi l^2} \right) ^{N/2} [det(\mathbf{A})]^{1/2} \ ,
\label{EQ:normalization}
\end{equation}
with $\textit{N}_x=\textit{N}_y=\textit{N}_z=\textit{N}^{1/3}$. 
The statistically averaged internal energy for one molecule consisting of $N$ monomers simplifies to
\begin{equation}
\left< \frac{E}{k_B T} \right> = N \int U e^{-\frac{1}{2} \mathbf{r}^T \mathbf{A'} \mathbf{r}} d\, r = N \int \frac{1}{2} \mathbf{r}^T \mathbf{A'} \mathbf{r} e^{-\frac{1}{2} \mathbf{r}^T \mathbf{A'} \mathbf{r}} \, .
\end{equation}
In one dimension,
\begin{eqnarray}
\left< \frac{E}{k_B T} \right>_x
& = &\frac{3 N_x}{2l^2} \int d\mathbf{x}\, \, \mathbf{x}^T \mathbf{A} \mathbf{x}\, e^{-\frac{3}{2l^2} \mathbf{x}^T \mathbf{A} \mathbf{x}}\nonumber \\
&=& N_x \frac{3}{2 l^2} \left[ \frac{N_x l^2}{3} \prod_{i=1}^{N_x} \sqrt{\frac{2 \pi l^2}{3 \lambda_i}} \right]=\frac{N}{2} \ ,
\end{eqnarray}
which gives, as the final result for the internal energy of one molecule consisting of $N$ monomers,
\begin{equation}
\left< \frac{E}{k_B T} \right> = \frac{3 N}{2} \  .
\end{equation}

\section*{APPENDIX II: THE DYNAMIC MEMORY FUNCITION}
\label{Appendix2}
We briefly report here the derivation of Eq.(\ref{EQ:MFMON}) starting from Eq.(\ref{cJ}). 
The product of the direct and projected forces is expressed as a function of the density field variables as
\begin{eqnarray}
\bigl< \mathbf{F}(0) \cdot \mathbf{F}^{\hat Q}(t)\bigr> \cong  \mathbf{F}(\mathbf{r}) \cdot \mathbf{F}(\mathbf{r}') 
\, \bigl<\rho_\alpha (\textbf{r};0)\rho_\gamma (\textbf{r};t)\bigr> \ .
\end{eqnarray}
Because the fluid is uniform and isotropic, the density fields can be replaced by their fluctuation variables, $\Delta \rho_\alpha (\mathbf{r},t)=\rho_\alpha (\mathbf{r},t)-\bigl<\rho_\alpha (\mathbf{r})\bigr>$, where the ensemble-averaged density field is approximated by $\bigl<\rho_\alpha (\mathbf{r})\bigr>\approx \rho g(r)$.
The correlation of the random forces is then expressed as
\begin{eqnarray}
\label{uhu}
\bigl< \mathbf{F}(0) \cdot \mathbf{F}^{\hat Q}(t)\bigr> &\cong& {\hat r} \cdot {\hat r'}  \rho^2 g(r)g(r') F(r) F(r') \nonumber \\
&&\times \frac{\bigl<\Delta\rho_\alpha (\mathbf{r})\Delta\rho_\gamma (\mathbf{r}',t)\bigr>}{\bigl<\rho_\alpha (r)\bigr>\bigl<\rho_\gamma (r')\bigr>} \ ,
\end{eqnarray} 
where we adopt a kind of "dynamical" Kirkwood superposition approximation in a weighted average form
\begin{eqnarray}
\label{aha}
\bigl<\Delta\rho_\alpha (\mathbf{r})\Delta\rho_\gamma (\mathbf{r}',t)\bigr> &\approx& \rho  \int d \mathbf{R} \ g(r) g(r') S(R,t)\nonumber \\
&&\times S(|\mathbf{r}- \mathbf{r}' + \mathbf{R}|t) \ .
\end{eqnarray} 
Eq.(\ref{aha}) describes the multipoint correlation between the density fluctuations at a distance $\mathbf{r}$ from segment $\alpha$ at time zero, and the density fluctuations a distance $\mathbf{r}'$ from segment $\gamma$ at time $t$. Because $\alpha$ and $\gamma$ can be on the same or on different polymer chains, no assumptions are made a priori about the relative importance of intra and intermolecular correlations. In this way, the chain connectivity does not play a dominant role in our description from the very beginning. We then calculate both intra and intermolecular contributions and show that both need to be included in the calculation of the memory function as intramolecular contributions are comparable in size to the intermolecular ones. 
Substitution of Eq.(\ref{aha}) into Eq.(\ref{uhu}) leads to Eq.(\ref{EQ:MFMON}).

\section*{ACKNOWLEDGEMENTS}
We acknowledge support from the National Science Foundation through grant DMR-0804145.
Computational resources were provided by LONI through the TeraGrid project supported by NSF. 
We thank James McCarty for his careful reading of the manuscript. 
We are grateful to G. S. Grest, V. G. Mavrantzas and coworkers for sharing the UA-MD computer simulation trajectories.


\end{document}